\input amstex
\documentstyle{amsppt}
%\magnification=1200

\loadbold
\define \hl{\;\hat\ll\;}
\define \pl{\ll^p}
\define \phl{\;\hat\ll^p\;}
\define \hpl{\;\widehat{\ll^p}\;}

\topmatter
\title
Universality of the Future Chronological Boundary
\endtitle
\author
Steven G. Harris
\endauthor
\address
Department of Mathematics, Saint Louis University,
St. Louis, MO 63103, USA\endaddress
\email harrissg\@slu.edu\endemail
\thanks
This research was supported in part by the
Department of Physics and Theoretical Physics of the
Australian National University, and in part by
NSF grant DMS9310477 to St. Louis University.
\endthanks
\abstract
The purpose of this note is to establish,
in a categorical manner, the universality of the
Geroch-Kronheimer-Penrose causal boundary when
considering the types of causal structures that may
profitably be put on any sort of boundary for a
spacetime.  Actually, this can only be done for the
future causal boundary (or the past causal boundary)
separately; furthermore, only the chronology
relation, not the causality relation, is considered, and the GKP
topology is eschewed.  The final result is that there is a unique
map, with the proper causal properties, from the future causal
boundary of a spacetime onto any ``reasonable"
boundary which supports some sort of chronological
structure and which purports to consist of a
future completion of the spacetime.  Furthermore, the
future causal boundary construction is categorically
unique in this regard.
\endabstract
\endtopmatter

\document

\head
Section I: Introduction and Discussion
\endhead

The causal boundary of Geroch, Kronheimer, and Penrose \cite{GKP}
(also explicated in \cite{HE}, section 6.8) has played a role in
helping to make clear the purely causal aspects of spacetimes,
especially at timelike or null infinity.  Being defined in terms
of the pasts and futures of points, it seems to be a ``natural"
object for focussing on such matters.  However, it has generally
been considered somewhat opaque in application and difficult
of calculation.  Is it still worth pursuing in spite of its
somewhat formidable reputation?  An affirmative answer may
be forthcoming if the intuitive sense of naturalness
accruing to the causal boundary can be justified in a
precise and general manner.  It is the purpose of this note
to do exactly that: to show that the causal boundary is a
``universal object" in the strict categorical sense, that is
to say, that in an appropriate mathematical category, the
causal boundary construction is the essentially unique way
of completing a spacetime in such a way as to yield a model
to which any other causal completion process can naturally
be compared (by means of a map from the causal boundary to
the other boundary).

However, some caveats are in order:  First of all, the categorical
constructions performed here operate only on the future causal
boundary (or, alternatively, only the past causal boundary), not
on the melding together of the future and past boundaries that
constitutes one of the more complicated issues in the full
Geroch-Kronheimer-Penrose (GKP) construction. 
Effectively, this is the portion of the GKP
construction which places an endpoint on each
future-endless causal curve in a spacetime,
leaving past endpoints unaccounted for.  Correspondingly, the
categorical comparison will be made only with any ``reasonable"
future completion of a spacetime---a completion of all future
causal curves only.  One way to understand why this is necessary
is by noting that although the future causal boundary completes all
future causal curves, and the past causal boundary completes all
past causal curves, the combination of the two creates additional
causal relations which can result in an object which is no longer
complete in either future or past.  (An example is given in the
appendix.)  

Second, in this presentation, only the chronology relation will
be addressed, not the causality relation; this is more a matter
of simplification than anything else:  Categorical results can be
obtained by working with only the one relation, so that is the
path followed here, though there is no barrier to including both
relations.  Accordingly, the boundary construction here is
called the future chronological boundary, allowing a distinction
from the GKP (future) causal boundary.

Finally, all topology is eschewed from the approach here, with
notions of convergence being treated by purely chronological
means.  This allows an enormous savings of complexity, as well
as the avoidance of one of the more difficult elements of the GKP
causal boundary.  (A topology can be inferred from the
chronological notion of convergence developed herein, but this
will not play an explicit role.) 

\bigpagebreak

As a model for the construction followed here, consider the
problem of completing a Riemannian manifold.  There is more than
one way to do this:  One could take the one-point
compactification, the Cauchy completion, or some arbitrary
isometric embedding into a complete metric space.  Part of the
issue is that one is liable to end up with a non-manifold as
one's complete object:  The category of Riemannian manifolds and
differentiable functions is not big enough to contain all the
completions we may wish to consider.  So the first thing to do is
to expand to a larger category, the category {\bf Met} of metric
spaces and continuous functions. We'll use Cauchy completion
as our method of completing not just any Riemannian manifold,
but any metric space at all.  This is actually a functor 
$\text{\bf C}: \text{\bf Met} \to  \text{\bf Met}_0$
from {\bf Met} to the subcategory of complete metric spaces,
{\bf Met$_0$}; this means that not only does each metric space
$X$ have a Cauchy completion $\text{\bf C}(X)$, but also that for
every map $f: X \to Y$ in {\bf Met}, there is an extension of $f$
to the completions, $\text{\bf C}(f): \text{\bf C}(X) \to
\text{\bf C}(Y)$, which works with compositions and with identity
maps, i.e., $\text{\bf C}(f \circ g) =
\text{\bf C}(f) \circ \text{\bf C}(g)$ and $\text{\bf
C}(\text{id}_X) = \text{id}_{\text{\bf C}(X)}$.  We need to take
note not only of the completion of each metric space $X$ but also
of its embedding into its completion, $i_X : X \to 
\text{\bf C}(X)$.  This is a natural transformation $\text{\bf
i} : \text{\bf I}_{\text{\bf Met}}\;\; \dot\to
\;\;\text{\bf U} \circ \text{\bf C}$, where {\bf I}$_{\text{\bf
Met}}$ is the identity functor on {\bf Met} and $\text{\bf U} :
\text{\bf Met}_0 \to \text{\bf Met}$ is the ``forgetful" (i.e.,
inclusion) functor; this means that for any map $f: X \to Y$, the
embeddings on $X$ and $Y$ work with the extension to $f$, i.e.,
$i_Y \circ f = \text{\bf C}(f) \circ i_X$ (also, as a
natural transformation, $i_X$ is considered as
technically mapping $X$ to $\text{\bf U}(\text{\bf C}(X))$).  The
importance of the Cauchy construction for completion is that it
is universal in the sense of being a left adjoint to the
forgetful functor; this means that for any map from a metric
space into a complete metric space, there is a unique extension
of that map to the Cauchy completion:  For any $X$ in {\bf Met},
any $Y$ in {\bf Met}$_0$, and any map $f : X \to \text{\bf U}(Y)$,
there is a unique map $f_0 : \text{\bf C}(X) \to Y$ such that
$\text{\bf U}(f_0) \circ i_X = f$ (more simply,
ignoring the forgetful functor:  for any $f: X \to Y$ with $Y$
complete, there is a unique $f_0 : \text{\bf C}(X) \to Y$ such
that $f_0 \circ i_X = f$).  Since left adjoints are
unique up to natural isomorphism (a natural transformation
consisting of isomorphisms) (\cite{M}, Corollary IV.1), this
tells us that Cauchy completion is {\it the\/} correct
construction for the ``free" completion of a metric space, in
exact analogy with the free vector space on a set:  It is
functorial, natural, and universal.  (But note that we had to
expand from Riemannian manifolds to the larger setting of metric
spaces in order to obtain a category in which the functoriality of
the construction comes into play.)
\medpagebreak

The general idea here is to define a category of objects which
includes both spacetimes and spacetimes that have boundaries
attached to them, and to show that the future chronological
boundary is a functorial, natural, and universal construction. 
The category, called chronological sets, consists, roughly
speaking, of objects that are spacetime-like in that they possess
a relation that works like the chronology relation in a
spacetime.  There is a great deal of freedom in choosing which
properties of the spacetime relation to abstract and axiomatize
for our new category.  The ones chosen here seem to be the
minimum ones needed to carry enough structure to be able to
replace topology as the means for defining convergence for
appropriate sequences of points.  The approach here is somewhat
reminiscent of the causal spaces axiomatized in
\cite{KP}, and some of the language (such as ``full" and
``chain") from that paper is borrowed for this one; but there is
a great deal of simplification achieved by concentrating solely
on the chronology relation. 

What is the practical impact of establishing the
universality of the future chronological
boundary?  How does this help in understanding
spacetimes?  The answer lies in being able
to relate any proposed future boundary for a
spacetime to what is essentially the GKP future
(pre-)boundary points, in a manner respecting the
chronological relations---so long as the proposed
boundary includes an extension of the chronology
relation of the spacetime and does so in a
``reasonable" manner (such as being
past-distinguishing).  Some examples at this point
may help:

First an artificial example:  Consider a spacetime $M$
obtained from Minkowski $2$-space ($\Bbb L^2$) by
excising a timelike line segment $L$ (such as $\{(0,t)
| 0 \le t \le 1\}$).  (A closely related example is
considered in detail at the end of section IV.)  The most
obvious way to put a helpful boundary on
$M$ is to embed it back into
$\Bbb L^2$ in the obvious way, yielding $L$ as the
boundary points for $M$.  However, this is not the only
useful possibility.  Let us expand our conception of
$M$ to include anything conformal to $\Bbb L^2 - L$. 
Then, considering that the metric may become
inherently singular as the points of $L$ are
approached, we may wish to choose a boundary
construction that respects $L$ as a physical blockage
between its left and right sides; typically, we may
choose as boundary two copies of $L$, one for the left
side and one for the right side of the slit in $M$,
joined at the top and the bottom, i.e., a boundary of
the form $\{(0^-,t) | 0 \le t \le 1\} \cup \{(0^+,t) | 0
\le t \le 1\}$, with $(0^-,0)$ and $(0^+,0)$
identified, as well as $(0^-,1)$ and $(0^+,1)$.
(By ``physical blockage" here I refer to the chronology
relation in the spacetime-{\it cum\/}-boundary:  We
have, for instance, $(-.1,.3) \ll (0^-,.5)$, and
$(.1,.3) \ll (0^+,.5)$, but no chronology relation
between $(-.1,.3)$ and $(0^+,.5)$.  This reflects an
impossibility of passage from one side of the slit to
the other.)  The GKP causal boundary produces precisely
this latter boundary, in addition to the standard
points at future and past null- and timelike-infinity
associated with Minkowski space.  The future
(pre-)boundary points of the GKP construction are
similar, with the only difference being that there is
a ``double-point" at the top of the boundary:
$(0^-,1)$ and $(0^+,1)$ are not identified but kept
distinct.  This last is the future chronological
boundary.

What, then, does the universality of the future
chronological boundary have to say in this example?  It
says that the future chronological boundary covers any
other (past-distinguishing) boundary.  For instance,
the full GKP causal boundary identifies the
double-point of the future chronological boundary into
a single point.  The first boundary mentioned,
consisting of $L$ itself, identifies all the pairs of
points $(0^-,t)$ and $(0^+,t)$.  Other boundaries can
be imagined, such as identifying $(0^-,t)$ and
$(0^+,t)$ for some but not all $t$.  (Note that
past-distinguishment in the boundary necessitates
retaining the identification of $(0^-,0)$ and
$(0^+,0)$.)  The point is that {\it any\/} appropriate
boundary construction for $M$, intending to add future
endpoints for all timelike curves, can be
mapped onto by the future chronological boundary, in a
manner that appropriately preserves the chronological
relation; that this mapping is unique; and that the
entire construction is categorical in nature. 

For a more realistic example, let our spacetime be
Schrwazschild space.  The standard picture
of the (future) singularity is of $\Bbb S^2 \times
\Bbb R^1$: a $2$-sphere (parametrized by $\phi$ and
$\theta$) at each point of $r = 0$ and any $t \in
\Bbb R$, using Schwarzschild coordinates.  We also
have future null-infinity ($\Im^+$ = $\Bbb S^2 \times
\Bbb R^1$: $r = \infty$, other coordinates arbitrary)
and future timelike-infinity ($i^+$, a single point). 
What other reasonable options are there for future
boundary points in Schwarzschild?  The boundary just
described is the future chronological boundary
(external Schwarzschild is conformal to a standard
static spacetime, which makes it easy to find its
causal boundary).  It follows that any
past-distinguishing way of putting a future boundary on
Schwarzschild must be a quotient of this:  We can
collapse the singularity to, for instance, a
projective space or lens space at each value of $t$,
and we can make identifications in $\Im^+$, but we
cannot blow up $i^+$ into a sphere or anything other
than a single point.

\bigpagebreak

It may be helpful to outline the some of the
complexities encountered in this presentation.  

The first step is to define a category which
contains both reasonable (strongly causal)
spacetimes and any such spacetime combined with a
boundary of any sort (including an extension of
the chronology relation to the boundary).  The
next step is to define a ``future-completion"
construction which will apply to any object in
this category; it should replicate the GKP future
(pre-)boundary construction when applied to
spacetimes, and should yield nothing new when
applied to an object which is already
``future-complete".  It is also intended that this
future-completion construction not change anything
in the original object, but just add additional
points. 

A problem arises in this, however:  The condition
used to define the the chronology relation for the
new (future boundary) points---the same criterion
as in the GKP construction---is not necessarily
compatible with the pre-existing chronology
relation in the original object.  (Among
spacetimes, global hyperbolicity prevents this
from happening, but the idea is to provide a
construction for all strongly causal spacetimes.) 
The problem is that if the same criterion is
applied to the original object, then more pairs of
points ought to be considered as chronologically
related than was originally the case.  For
instance, if the object is the spacetime $M$ from
the example above, then the criterion applied to
$M$ would have $(-1,-1) \ll (.5,1)$; but
there is no timelike curve between those points in
$M$.  Among other things, this incompatibility
interferes with the functoriality of the
future-completion construction.

The solution is to define an additional
construction (``past-determination") on these
objects whose sole purpose is to add in
precisely those additional chronology relations
so that when the future-completion
construction is applied, the result is compatible
with the chronology relation in the past-determined
object.  This defeats the intent of preserving the
original chronology relation for the object, but
there is no getting around this point.  Indeed,
this is precisely the tack chosen by GKP:  The
chronology relation in the GKP construction
results in additional relations within the
spacetime, exactly the same ones as the
past-determination construction here.  The
exposition here essentially separates out this
portion of the GKP construction for its own
categorical and universal explication.

One still has to contend with the question of
whether it makes a difference if one first
past-determines and then future-completes, or does
it the other way around.  As it turns out, these
operations are categorically commutative; but that
takes some doing to establish.

Finally, one needs to make a comparison between
the future chronological boundary construction
here---the GKP future (pre-)boundary---and the
full GKP causal boundary.  A difficulty arises in
that the simple definition of future-complete used
here is not generally satisfied by the GKP causal
boundary; a slight generalization is needed,
allowing, for instance, for the identification of
points such as in the cuasl boundary of the example
$M$ above.  But all goes through, as exemplified in
that example.

In summation: Section II introduces the basic
category  and the future-comple\-tion operation;
this applies, without further modification, to
globally hyperbolic spacetimes.  Section III
defines the past-determination operation, extending
the results of Section 1 to, for instance, all
strongly causal spacetimes.  Section IV addresses
the generalization that is necessary in order to
include such completions as those deriving from
the full GKP construction. 

\bigpagebreak
A word on what this paper addresses and what it
does not:

The intent here is to provide a justification for
the use of the causal boundary construction.  This
is accomplished largely though showing how at
least the future portion of that construction
provides a universal standard of comparison for
any other future-completing boundary, in tems of
the chronology relation.

One obvious failing of this enterprise is to
evince a similar standard of comparison for the
full causal boundary.  This, however, I believe to
be innately impossible:  As evidenced in the
Appendix, the full causal boundary simply does not
result in a complete object.  This does not
necessarily mean that the full causal boundary is
an inappropriate tool, but it does detract from
hopes of its having a universal role in boundary
considerations.

Another element missing here is the causality
relation.  I believe it possible to do much of the
same work here with the causality relation
included, but it seems an unnecessary complication
at this point.  The chronology relation suffices
to cover much of the necessary structures, and it
yields a satisfactory universality result.

Topology is probably the most complex element
missing from this development.  It would be
desireable to have a universality result which
included full topological information, not just
chronological.  This is a matter under
investigation and the subject for another paper.

\head
Section II: Chronological Sets and Future Completion
\endhead

The intent here is to abstract and axiomatize the
chronology relation in a spacetime:  A point $x$
chronologically precedes a point $y$,
written $x \ll y$, if and only if there is a
future-directed timelike curve from $x$ to $y$. 
This is the relation that must be
emulated for the new category.

We start with a subset of the needed properties:
\smallpagebreak

A {\it pre-chronological set\/} is a set
$X$ together with a relation $\ll$ satisfying the
following conditions:
\roster
\item $\ll$ is transitive: $x \ll y \ll z$
implies $x \ll z$
\item $\ll$ is non-reflexive: for all $x$,
$\slanted{not}(x \ll x)$ 
\item $X$ has no isolates: for all $x$,
$I^-(x) =\{y | y \ll x\}$, the {\it past\/} of $x$,
or $I^+(x) =\{y | x \ll y\}$, the {\it future\/} of
$x$, is non-empty
\item $\ll$ is full: for any $x \ll y$, there is
some $z$ with $x \ll z \ll y$
\endroster

For any non-empty subset $A$ of a 
pre-chronological set $X$, the {\it past\/} of
$A$, $I^-[A]$, is $\bigcup_{a \in A}I^-(a)$
(the {\it future\/} of $A$, and all subsequent
dualizations, are defined analogously).  A non-empty
subset $P$ is a {\it past set\/} if
$I^-[P] = P$.  (In general, $(\;)$ will be used to
indicate application of an operator to an entity
to be regarded as a single-element argument,
$[\;]$ will be used to indicate application to
each of the various elements making up the
argument.)

\proclaim{1. Proposition} In a 
pre-chronological set, any non-empty set of the
form $I^-[A]$ is a past set.\endproclaim

\demo{Proof} By transitivity, $I^-[I^-[A]]
\subset I^-[A]$. Now suppose $x \ll a$ for $a \in
A$; then, by fullness, there is some $y$ with $x
\ll y \ll a$.  Thus, $I^-[I^-[A]] = I^-[A]$.
\qed\enddemo

A past set is {\it decomposable\/} if it is the
union of two proper subsets that are past sets. 
An indecomposable past set is often called an
{\eightpoint IP}. 

\proclaim{2. Theorem} In a  pre-chrono\-logical
set, a past set P is indecomposable if and only if
for every $x$ and $y$ in $P$, there is some $z \in
P$ with $x \ll z$ and $y \ll z$.\endproclaim

\demo{Proof} (This theorem is due to Geroch, Kronheimer, and
Penrose; the proof here is adapted from
\cite{HE}, in the proof of Proposition 6.8.1.) 

Suppose $P$ is decomposable as
$P_1 \cup P_2$ (proper past subsets).  Then neither
$P_1 \subset P_2$ nor $P_2 \subset P_1$, so we
can find $x_1 \in P_1 - P_2$ and $x_2 \in P_2 -
P_1$.  Then there can be no $y \in P$ with $x_1
\ll y$ and $x_2 \ll y$:  If $y$ is in, say,
$P_1$, then since $x_2 \ll y$, $x_2 \in P_1$
also, which is false.

Suppose $P$ is indecomposable.  For any $x_1$ and
$x_2$ in $P$, let $Q_i = \{q \in P | \mathbreak
\slanted{not}(x_i \ll q)\}$, and let
$P_i = I^-[Q_i]$ (for each $i$).  Note that $P_i
\subset Q_i$:  For $x \in P_i$, there is some $q
\in P$ with $x \ll q$ and $\slanted{not}(x_i \ll
q)$; thus, $x \in P$ and $\slanted{not}(x_i \ll x)$
(lest $x_i \ll q$), so $x \in Q_i$.  Therefore,
$P_i$ cannot be all of $P$, since $Q_i$ is not
(since $x_i$ is in $P$ and $P = I^-[P]$, there
must be some $p_i \in P$ with $x_i \ll p_i$). 
Since $P_1$ and $P_2$ are both past sets, it
follows from $P$ being an {\eightpoint IP} that $P \ne P_1 \cup
P_2$, i.e., there is some $p \in P$ with $p
\not\in I^-[Q_1]$ and $p \not\in  I^-[Q_2]$. 
Now, since $P = I^-[P]$, there is some
$z \in P$ with $p \ll z$; it then follows that
$z$ must be in neither $Q_i$, i.e., that $x_1 \ll
z$ and $x_2 \ll z$.\qed\enddemo

A {\it future chain\/} in a 
pre-chronological set $X$ is a sequence $\{x_n\ | n
\ge 1\}$ of points in $X$ such that for all $n$,
$x_n \ll x_{n+1}$.

A subset $S$ of a pre-chronological set
$X$ is called {\it dense\/} in $X$ if for all $x$
and $y$ in $X$ with $x \ll y$, there is some $s$
in $S$ with $x \ll s \ll y$.  $X$ is {\it
separable\/} (in a chronological sense) if there is a countable
dense set.

The following theorem shows how future chains in a
chronological set replace future timelike curves in a spacetime;
this is adapted from Proposition 6.8.1 in [HE], due to
Geroch, Kronheimer, and Penrose, which makes essentially the same
statement with ``timelike curve" in the place of ``chain".

\proclaim{3. Theorem} A past set in a separable
pre-chronological set is indecomposable  if
and only if it is the past of a future chain.
\endproclaim

\demo{Proof}  Let $P$ be $I^-[\{x_n | n \ge 1\}]$
for $\{x_n\}$ a future chain; then for any $p_1$
and $p_2$ in $P$, there are numbers $n_1$ and
$n_2$ with $p_1 \ll x_{n_1}$ and $p_2 \ll
x_{n_2}$.  Let $k = 1 + \max\{n_1,n_2\}$; then
$x_k$ is in $P$ and has each $p_i$ in its past.

Let $P$ be an {\eightpoint IP}. If $S$ is the countable
dense set for the pre-chronological set,
then $P \cap S = \{p_n | n \ge 1\}$ is a dense set
in $P$.  Pick $x_1 \ll p_1$ (one can always
arrange to have the first element of $S$ have a
non-empty past).  Suppose we've defined
$\{x_1, ..., x_k\}$ with $x_i \in P$ and
$x_{i-1} \ll x_i$ and $p_i \ll x_i$ for each $i
\le k$.  Then by Theorem 2 there is some
$x_{k+1} \in P$ with $x_k \ll x_{k+1}$ and
$p_{k+1} \ll x_{k+1}$.  This defines`for us the
future chain $\{x_n\}$ in $P$ with $p_n \ll
x_n$ for each $n$.  Clearly, $I^-[\{x_n\}] \subset
P$.  For any $x \in P$, since $P$ is a past set,
there is some $y \in P$ with $x \ll y$; then there
is some $k$ with $x \ll p_k \ll y$, so $p_k \ll
x_k$ shows us $x \ll x_k$.  Thus, $I^-[\{x_n\}] =
P$. \qed\enddemo

We will say that an {\eightpoint IP} $P$ is {\it generated\/}
by any future chain $c$ for which $P = I^-[c]$.

A pre-chronological set is called {\it
past-distinguishing\/} if for points $x$ and $y$,
$I^-(x) = I^-(y)$ implies $x = y$.  (Unlike nearly all the other
terms defined here, this one is traditional, following the usage
in \cite{HE}.)

For a future chain $c = \{x_n | n \ge 1\}$, a
point $x$ is called a {\it future limit\/} of
$c$ if $I^-(x) = I^-[c]$.   Note that if $x$ is
a future limit of $c$, then for all $n$, $x_n \ll
x$, since $I^-(x)$ includes $I^-(x_{n+1})$. 
Also, if $x$ is a future limit of $c$, then it is
also a future limit of every sub-chain of $c$;
and, conversely, if $x$ is a future limit of some
sub-chain of $c$, then it is also a future limit
of $c$.  Clearly, in a past-distinguishing
pre-chronological set, a future chain can have at
most one future limit.  (We could define a topology from
this notion of convergence---a set is (future-) closed
if it contains all future limits of future chains within
it---but this will not be pursued here.)

A pre-chronological set that is separable
is called a {\it chronological set\/}.  A 
chronological set for which every future chain has
a future limit is called {\it future-complete\/}.

It will prove useful to have the $\ll$ relation
determined, in a certain sense, by $I^-$:
Specifically, call  a pre-chronological set
$X$ {\it past-determined\/} if for any $x$ and $y$
in $X$, if $I^-(x) $ is non-empty and for some $w
\ll y$, $I^-(x) \subset I^-(w)$, then $x \ll y$.
(We can also say that $x$ is past-determined if
$I^-(x)$ is non-empty and whenever $I^-(x) \subset
I^-(w)$, $x \ll y$ for all $y \gg w$.)

\proclaim{4. Theorem} A strongly causal spacetime
is a past-distinguishing chronological set; if it
has the property that the past of $x$ is
contained in the past of $y$ implies that $x$
causally precedes $y$ (for instance, any globally
hyperbolic spacetime), then it is past-determined. 
A future limit of a future chain in a spacetime is
the same as a topological limit.
\endproclaim
 
\demo{Proof} A spacetime is clearly a
pre-chronological set; it is separable in the
chronological sense because it is separable in the
topological sense.  A strongly causal spacetime is
past-distinguishing (see, e.g., \cite{HE}). 

If $I^-(x) \subset I^-(w)$ implies $x \in
J^-(w)$ (points causally preceding $w$), then with
$w \ll y$ also, we have $x \ll y$.  Globally
hyperbolic spacetimes are examples of this:  For
$I^-(x) \subset I^-(w)$, consider any $z \ll x$;
then since $z \in I^-(x)$, $z$ is also in $I^-(w)$,
so $z \ll w$.  It follows that the space $C(z,w)$ of
causal curves from $z$ to $w$ is compact in the $C^0$
topology.  Let $\sigma$ be a future timelike curve
with $\sigma(0) = z$ and $\sigma(1) = x$; for all
$t <1$, there is also a future timelike curve
$\tau_t$ from $\sigma(t)$ to $w$.  Let $c_t$ be the
concatenation of $\sigma|_{[0,t]}$ with $\tau_t$;
then the family of causal curves $\{c_t\}$ has a
causal limit curve $c_1$ in $C(z,w)$; since $c_1$
must pass through $x$, this shows $x \in J^-(w)$.

In a strongly causal spacetime, each
point has a fundamental neighborhood system
$\{U_n\}$ such that for each $n$, $U_n$ is
geodesically convex and for any point
$x \in U_n$, any other point $y \in U_n$ is
timelike-related to $x$ iff it is joined to $x$
by a timelike geodesic in $U_n$ ($x \ll y$ but
not joined by a timelike geodesic in $U_n$ implies
that the timelike geodesic connecting them must
exit and re-enter $U_n$, which can be prevented
with correct choice of $U_n$).  Let
$c = \{x_n\}$ be a future chain.  First suppose
that $c$ has a topological limit $x$; then $c$
eventually enters all such neighborhoods $U_n$ of
$x$ as just described.  For any $y \ll x$, we
have $x \in I^+(y)$, a neighborhood of $x$; thus,
eventually, $x_n \in I^+(y)$, i.e., $y \ll x_n$:
$I^-(x) \subset I^-[c]$.  If any
$x_m$ fails to be in the past of $x$, the same is
true for all $x_k$ with $k > m$.  Consider such an
$x_k$ in a neighborhood $U_n$ of $x$ as above;
due to the chronological relation in $U_n$ being
given by the geodesic structure, this is
essentially the same causal relations as exist in
Minkowski space:  In particular, $x_{k+1}$ is not
in $\slanted{closure}(I^-(x))$, and all $x_j$ for $j
> k+1$ are in $I^+(x_{k+1})$, whose closure does
not include $x$.  This violates $x$ being the
limit of $c$; thus, we must have $I^-(x) =
I^-[c]$.  Conversely, suppose that $I^-(x) =
I^-[c]$; then we have $x_n \ll x$ for each $n$
(since $x_n \in I^-[c]$).  Using the
neighborhoods $U_n$ of $x$ from above, we can
find a past chain $\{y_i\}$ in $I^+(x)$ and a
future chain $\{z_i\}$ in $I^-(x)$ with $x$ the
topological limit of both chains and the sets
$\{V_i = I^-(y_i) \cap I^+(z_i)\}$ a fundamental
neighborhood system for $x$.  Then for each $i$,
since $z_i \in I^-(x) = I^-[c]$, $x_n \gg z_i$
for sufficiently large $n$.  Since $x_n \ll x
\ll y_i$, this gives $x_n \in V_i$ for
sufficiently large $n$, i.e., $x$ is the
topological limit of $c$.
\qed\enddemo

For a chronological set $X$, let
$\hat\partial (X)= \{P | P \text{ is an {\eightpoint IP} in } X
\text{ such that } P \text{ is not}\mathbreak
I^-(x) \text{ for any point } x \in
X\}$, the {\it future chronological boundary\/} of
$X$. Let $\hat X = X \cup \hat\partial(X)$, the
{\it future chronological completion\/} of $X$. 
Define an extension of $\ll$ on $X$ to $\hat X$
by the following:
\roster
\item among points in $X$, there is no change in
$\ll$
\item for $x \in X$ and $P \in \hat\partial(X)$,
$x \ll P$ iff $x \in P$
\item for $x \in X$ and $P \in \hat\partial(X)$,
$P \ll x$ iff for some $y \in X$ with $y \ll x$,
$P \subset I^-(y)$
\item for $P \in \hat\partial(X)$ and $Q \in
\hat\partial(X)$, $P \ll Q$ iff for some $y \in Q$,
$P \subset I^-(y)$.
\endroster
Where it is necessary to distinguish between the
relation on $X$ and that on $\hat X$, $\hat I^-$
will be used for the latter. (Thus, $x \in \hat
I^-(y)$ iff $x \in I^-(y)$, and so on.)  Note
that $\hat I^-(P) = P \cup \{Q \in \hat\partial(X)
| Q \subset I^-(w) \text{ for some } w \in P\} =
\hat I^-[P]$ (the first being the past of a
single point in $\hat X$, the last being the
past of a subset of $\hat X$); in particular, $\hat I^-(P) \cap X
= P$.  Also note that
$P$ is a future limit of a future chain $c$ (in
$\hat X$) if and only if $P$ is generated by $c$
($I^-[c] = P$ iff $I^-[c] = I^-[P]$ iff $\hat
I^-[c] = \hat I^-[P]$ iff $\hat I^-[c] = \hat
I^-(P)$).

\proclaim{5. Theorem} Let $X$ be a chronological set.  Then 
\roster
\item $\hat X$ is a chronological set;
\item if $X$ is past-distinguishing, then so is $\hat X$;
\item if $X$ is past-determined, then so is $\hat X$; and
\item $\hat X$ is future-complete.
\endroster
\endproclaim

\demo{Proof} (1) $\hat X$ is a chronological set:

This is a matter of routine checking of various cases. 
For instance, to show transitivity, one of the cases to
be established is that for $x$ in $X$ and $P$ and $Q$
in $\hat\partial (X)$, if $x \ll P \ll Q$, then $x \ll
Q$:  We have $x \in Q$ and $Q \subset I^-(w)$ for some
$w \in R$; then $x \ll w$, and, since $R$ is a past
set, $x \in R$, i.e. $x \ll R$.

A set which is dense in $X$ is dense in $\hat X$ as
well (routine to establish).

Note that $\hat X$ has no isolates,
since for any
$P \in \hat\partial(X)$, $\hat I^-(P)$ includes
the elements of $P$, which is
non-empty.\smallpagebreak

(2) If $X$ is past-distinguishing, then  
$\hat X$ is also ($x$, $P$, and $Q$ as above, with $y
\in X$):

Suppose $\hat I^-(x) = \hat I^-(y)$.  Then, in
particular, $I^-(x) = I^-(y)$, so $x = y$.

Suppose $\hat I^-(x) = \hat I^-(P)$.  Then, in
particular, $\hat I^-(P) \cap X = I^-(x)$;
but this means $P = I^-(x)$, which is forbidden by
the definition of $\hat\partial(X)$.

Suppose $\hat I^-(P) = \hat I^-(Q)$.  Then, in
particular, $\hat I^-(P) \cap X = \hat I^-(Q)
\cap X$, which just means precisely that $P = Q$.
\smallpagebreak

(3) If $X$ is past-determined, then 
$\hat X$ is also (notation as above):

First note
that if $\hat I^-(x)$ is non-empty, then so is
$I^-(x)$ ($P \ll x$ requires some $w \ll x$), and
$\hat I^-(P)$ is always non-empty.

Suppose $\hat I^-(x) \subset \hat I^-(w)$ with $w
\ll y$; then $I^-(x) \subset I^-(w)$, so $x \ll
y$.  Suppose $\hat I^-(x) \subset \hat I^-(Q)$
with $Q \ll y$, i.e., $Q \subset I^-(w)$ and $w \ll
y$; then $I^-(x) \subset Q$, so $I^-(x) \subset
I^-(w)$, so $x \ll y$.

Suppose $\hat I^-(x) \subset \hat I^-(w)$ with $w
\in R$ and $\hat I^-(x)$ non-empty; then
we can find
$z \in R$ with $w \ll z$, so the previous result
gives us $x \ll z$, so $x \in R$, so $x \ll R$. 
Suppose $\hat I^-(x)
\subset \hat I^-(Q)$ with $Q \subset I^-(w)$ and
$w \in R$; then pick $z$ the same, and the
previous result gives $x \ll z$, so $x \in R$,
i.e., $x \ll R$.

Suppose $\hat I^-(P) \subset \hat I^-(w)$ with $w
\ll y$; then $P \subset I^-(w)$, so $P \ll y$. 
Suppose $\hat I^-(P) \subset \hat I^-(Q)$ with $Q
\subset I^-(w)$ and $w \ll y$; then $P \subset
Q$, so $P \subset I^-(w)$, so $P \ll y$.

Suppose $\hat I^-(P) \subset \hat I^-(w)$ with $w
\in R$; then pick $z \in R$ (i.e., $z \ll R$) with
$w \ll z$, and the previous result yields $P \ll
z$, so $P \ll R$.  Suppose $\hat I^-(P) \subset
\hat I^-(Q)$ with $Q \subset I^-(w)$ and $w \in R$
(i.e., $w \ll R$); then the previous result yields
$P \ll w$, so $P \ll R$.\smallpagebreak

(4) $\hat X$ is future-complete:

First consider a future chain $c =
\{x_n\}$ made up of points of $X$.  Let $P =
I^-[c]$; then $P$ is an {\eightpoint IP} in $X$.  If there is a
point $x \in X$ with $I^-(x) = P$, then $x$ is a
future limit of $c$:  Since each $x_n \ll x$, for
any $\alpha \in \hat I^-(x_n)$, $\alpha \ll x$
also; for $y \in I^-(x)$, we have $y \in I^-[c]$;
and for $Q \ll x$ with $Q \in \hat\partial(X)$, we
have $Q \subset I^-(w)$ for some $w \ll x$, hence,
$w \ll x_n$ for some $n$, so $Q \ll x_n$.  Now
suppose, on the other hand, that $P \in
\hat\partial(X)$; then $P$ is a future limit of
$c$, as noted previously.

Now consider a future chain $c$ in which there is
some sub-chain $c'$ consisting of points in $X$. 
By the paragraph above, $c'$ has a future limit,
which is then also a future limit for $c$.

For a future chain consisting solely of elements
of $\hat\partial X$, just interpolate elements of
$S$ (the countable dense subset of $X$) into the
chain, resulting in a new future chain to which
the previous paragraph applies; this has the same
future limits as the original one.
\qed\enddemo   

Let $f: X \to Y$ be a set map between
pre-chronological sets.  We will call $f$
{\it chronological\/} if $x \ll y$ in $X$ implies
$f(x) \ll f(y)$ in $Y$; $f$ will be called {\it
future-continuous\/} if, in addition, whenever
$x$ is a future limit of a future chain $c$ in
$X$, $f(x)$ is a future limit of $f[c]$
(necessarily a future chain in $Y$).  For any
chronological set $X$, we have the
future-continuous map $\hat\iota_X: X \to \hat X$
given by $\hat\iota_X(x) = x$; this is the {\it
standard future injection\/} of $X$ (that it is
future-continuous follows directly from the
definitions).

For $M$ and $N$ strongly causal spacetimes, a function
$f:M \to N$ is both past- and future-continuous if and
only if it is (topologically) continuous and carries
timelike curves to timelike curves, preserving
time-orientation.  This can be seen by examining, for
a point $x \in M$ with sequence $\{x_n\}$ topologically
converging to $x$, a fundamental neighborhood system of
the form described in the proof of Theorem 4,
$\{V_i = I^-(y_i) \cap I^+(z_i)\}$ for $\{y_i\}$ a past
chain and $\{z_i\}$ a future chain, both with $x$ as
topological limit.  With
$f$ both past- and future-continuous, we have
$f(x)$ is the topological limit of the chains
$\{f(y_i)\}$ and $\{f(z_i)\}$, so $\{W_i = I^-(f(y_i))
\cap I^+f(z_i))\}$ is a fundamental neighborhood system
for $f(x)$.  Eventually, $x_n \in V_i$, whence $f(x_n)
\in W_i$; thus, $f$ is topologically continuous. 
However, a merely future-continuous function need not be
continuous:  Consider $f: \Bbb L^2 \to \Bbb L^2$ defined
by $f(x,t) = (x,t)$ for $t > 0$ and $f(x,t) = (x,t-1)$
for $t \le 0$ ($\Bbb L^n$ denotes Minkowski $n$-space).

Let $f: X \to Y$ be a chronological map between
chronological sets with $Y$ past-distinguishing; we
will define a map $\hat f: \hat X \to \hat Y$, the
{\it future completion\/} of $f$.  First we define
$\bar f : \Cal P(X) \to \Cal P(Y)$ (the power sets)
by $\bar f(S)$ = $I^-[f[S]]$.  For any point $x \in
X$, we just let $\hat f(x) = f(x)$.  For an {\eightpoint IP} $P
\in \hat\partial(X)$, first note that $\bar f(P)$
is an {\eightpoint IP} in $Y$:  This is because for any future
chain $c$ generating $P$, $\bar f(P) =
I^-[f[c]]$.  If there is some $y \in Y$ with
$I^-(y) = \bar f(P)$, then we let $\hat f(P) =
y$; otherwise, we let $\hat f(P) = \bar f(P)$.
(Note that $\hat f$ can be defined even if $Y$ is
not past-distinguishing, so long as there is
never more than a single point $y \in Y$ for
which $I^-(y)$ is the same as a given $\bar
f(P)$.)

\proclaim{6. Proposition} Let $f : X \to Y$ be a
chronological map between chronological sets
with $Y$ past-determined and past-distinguishing
(or, more generally: for any future chain in $X$ with
no future limit in $X$, its image under
$f$ does not have more than one future limit in
$\hat Y$);  then $\hat f$ is also chronological. 
If $f$ is future-continuous, then $\hat f$ is the 
unique future-continuous map such that $\hat f
\circ \hat\iota_X = \hat\iota_Y \circ f$.\endproclaim

\demo{Proof} That  $\hat f \circ
\hat\iota_X = \hat\iota_Y \circ f$ is clear.  From
this and $f$ being chronological it follows that
$x \ll y$ in $X$ implies $\hat f(x) \ll \hat
f(y)$.  Suppose $x \ll P$ for $x \in X$ and $P \in
\hat\partial(X)$.  Let $P$ be generated by a
future chain $c = \{x_n\}$, so that $f[c]$
generates $\bar f(P)$.  We have $x \in P$, so $x
\ll x_n$ for some $n$; thus, $f(x) \ll f(x_n)$,
so $f(x) \in \bar f(P)$.  If $\bar f(P) \in \hat
\partial(Y)$, this is all we need for $\hat f(x)
\ll \hat f(P)$; otherwise, for some $y \in Y$,
$\bar f(P) = I^-(y)$, so $f(x) \in I^-(y)$, orM$f(x) \ll y$, so $\hat f(x) \ll \hat f(P)$.  If $P
\ll x$ (same notation), then
$P \subset I^-(w)$ for some $w \ll x$.  With the future
chain $c$ as before, we have for all $n$, $x_n
\ll w \ll x$, so $f(x_n) \ll f(w) \ll f(x)$. 
This gives us $\bar f(P) \subset I^-(f(w))$.  If
$\bar f(P) \in \hat\partial(Y)$, then $\hat f(P)
\ll \hat f(x)$; otherwise, for some $y \in Y$,
$\bar f(P) = I^-(y)$, so $y \ll f(x)$ by $Y$ being
past-determined.  Finally, suppose
$P \ll Q$ for $P$ and $Q$ in $\hat\partial(X)$;
then there is some $x \in X$ with $P \ll x \ll Q$,
so the previous results apply.

Suppose $f$ is future-continuous.  Clearly, for a
point $x \in X$ being a future limit of a future chain
of points $c = \{x_n\}$ in $X$, the same is true
for $\hat f(x)$ and $\{\hat f(x_n)\}$.  For $P \in
\hat\partial(X)$ being a future limit of the
chain, since $P$ is generated by $c$, $\bar
f(P)$ is generated by $f[c]$.  Thus, if  $\bar
f(P) \in \hat\partial(Y)$, then it is the future
limit of $f[c]$, and we are done, since $\hat
f(P) = \bar f(P)$.  Otherwise, for some $y \in
Y$, $\bar f(P) = I^-(y)$ and again we are done,
since $\hat f(P) = y$ and $y$ is a future limit of
$f[c] = \hat f[c]$.  We need concern ourselves
solely with chains of points in $X$, for any
other sort can have points of $X$ interpolated
into it with no change in future limits.

The restriction of $\hat f$ to $X$ is
uniquely determined by $\hat f \circ \hat\iota_X =
\hat\iota_Y \circ f$.  For $\hat f$ to be
future-continuous, for any $P \in
\hat\partial(X)$, since $P$ is a future limit of
any chain $c$ generating it, we must have $\hat
f(P)$ a future limit of $f[c]$.  With $Y$
past-distinguishing (or, more generally, with the
definition of $\hat f$ above yielding a
well-defined function), this is unique.\qed\enddemo

Another way to express the previous result is more closely related
to adjunctions:

\proclaim{7. Corollary} Let $f : X \to Y$ be a
future-continuous function between chronological
sets with $Y$ past-determined,
past-distinguishing, and future-complete (or, more
generally: for any future chain in $X$, its image
under $f$ has exactly one future limit in $Y$,
which is past-determined); then there is a unique
future-continuous function
$\hat f : \hat X\to Y$ such that $\hat f \circ \hat\iota_X
= f$.\endproclaim

\demo{Proof} We need but note that with $Y$ being future complete,
$\hat Y = Y$ and $\hat\iota_Y$ is the identity map on $Y$; then
Proposition 6 applies.  
\qed\enddemo

Conversely, we could have started with Corollary 7 and deduced
Proposition 6:  Given future-continuous $f: X \to Y$ with $Y$
past-determined and past-distinguish\-ing, we have $\hat\iota_Y
\circ f: X \to \hat Y$ fulfilling the hypotheses of Corollary 7,
yielding $\hat f: \hat X \to \hat Y$ as the unique
future-continuous map such that $\hat f \circ \hat\iota_X = 
\hat\iota_Y \circ f$.
\smallpagebreak

These results allow us to define some
categories, some functors, a natural transformation,
and an adjunction:  Let {\bf Chron} be the
category of chronological sets with morphisms all
chronological functions; since the composition of
chronological functions is clearly chronological,
this is a category.  Let {\bf PdetPdisChron} be the
subcategory of past-determined,
past-distinguishing chronological sets, with
morphisms all future-continuous functions; again,
a composition of future-continuous functions is
future-continuous, so this is a category.  Let
{\bf FcplPdetPdisChron} be the full subcategory of
{\bf PdetPdisChron} of objects which are also
future-complete.  Then future completion is a
functor $\;\widehat{}\; : \text{\bf PdetPdisChron}
\to \text{\bf FcplPdetPdisChron}$ (we need only check that
$\widehat{g \circ f} = \hat g \circ \hat f$; this
follows from the uniqueness part of Proposition 6,
since $\hat g \circ \hat f$ has the requisite
properties); while Theorem 5 gives us the right
properties for the future completion of any chronological set,
Proposition 6 gives us functoriality only for this subcategory.  We
also have the forgetful functor
$\hat{\text{\bf U}} : \text{\bf FcplPdetPdisChron} \to
\text{\bf PdetPdisChron}$.  
The maps $\hat\iota_X$ define a natural
transformation $\hat{\boldsymbol\iota} :
\text{\bf I}\;
\dot\to \;\;\hat{\text{\bf U}} \circ \;\widehat{}\;$
(where
$\text{\bf I}$ is the identity functor on {\bf PdetPdisChron}). 
Finally, Proposition 7 gives us the universality
condition on $\hat{\boldsymbol\iota}$ making
$\;\widehat{}\;$ left adjoint to $\hat{\text{\bf U}}$. 
From category theory we obtain

\proclaim{8. Theorem} The future completion
functor and the standard future injection are, up
to natural isomorphism, the unique construction
yielding a future-continuous map
from any past-determined, past-distinguishing
chronological set to a future-complete,
past-determined, past-distinguish\-ing
chronological set, consistent with all
future-continuous maps and yielding a unique
extension for any future-continuous map into a
future-complete, past-determined,
past-distin\-guish\-ing chronological 
set. \linebreak\qed\endproclaim

To restate the previous three theorems in summary form:

\roster

\item For $X$ in {\bf Chron}, $Y$ in {\bf PdetPdisChron}, and $f:
X \to Y$ future-continuous, $\hat f: \hat X \to \hat Y$ is a
future-continuous map with $\hat f
\circ \hat\iota_X = \hat\iota_Y \circ f$.

\item For $X$ in {\bf Chron}, $Y$ in {\bf FcplPdetPdisChron}, and
$f: X \to Y$ future-continuous, $\hat f : \hat X \to Y$ is the
unique future-continuous map with $\hat f \circ \hat\iota_X
= f$.

\item Restricted to {\bf PdetPdisChron}, the construction of $\hat
X$, $\hat f$, and  $\hat\iota_X$ is functorial, natural, and
universal, hence, unique as a way to fulfill (1) and (2) in such a
manner.
\endroster

\head
Section III: Past-Determination Functor
\endhead

It is quite common for a spacetime not to be
past-determined, so there is a limited
application of the functor $\;\widehat{}\;$ to
spacetimes.  However, this can be remedied if we
are willing to create new $\ll$ relations in the
spacetime, in order to make it past-determined. 
We will do this generally:  

Consider a pre-chronological set $X$; using the same set $X$, let
us define a new relation, $\ll^{p}$:  For any $x$ and $y$ in $X$,
set $x \ll^{p} y$ if  $x \ll y$ or if
$I^-(x)$ is non-empty and for some
$w \ll y$, $I^-(x) \subset I^-(w)$; we'll use
$I^{-p}$ to denote the past of a point (or set)
using the $\ll^{p}$ relation.  Clearly, 
$\ll^{p}$ is an extension of $\ll$.  Note that if
$X$ is past-determined, then $\ll^{p}$ is identical
to $\ll$. 

Two properties of $\pl$ should be noted:  First, for any
$x$, $I^-[I^{-p}(x)] = I^-(x)$: If $y \ll z
\ll^{p} x$, then $z \ll x$ (and so $y \ll x$) or $I^-(z) \subset
I^-(w)$ for some
$w \ll x$; we have $y \in I^-(z)$, so we get $y \in
I^-(w)$, i.e., $y \ll w$, so $y \ll x$.  On the
other hand, if $y \ll x$, then by fullness there is
some $z$ with $y \ll z \ll x$; this gives us $z
\ll^{p} x$, so $y \in I^-[I^{-p}(x)]$.  Second, $I^-(x)
\subset I^-(y)$ if and only if $I^{-p}(x) \subset
I^{-p}(y)$:  Given $I^-(x) \subset
I^-(y)$,  if $z \pl x$, then we have $z \ll x$ (so
$z \in I^-(x) \subset I^-(y) \subset I^{-p}(y)$) or we
have $I^-(z) \neq \emptyset$ and
$I^-(z) \subset I^-(w)$ for some $w \ll x$; in the latter case,
we have $w \ll y$ also, so $z \pl y$.  On the other hand,
given $I^{-p}(x) \subset I^{-p}(y)$, we obtain, from
the first property noted,  $I^-(x) = I^-[I^{-p}(x)]
\subset I^-[I^{-p}(y)] = I^-(y)$.  In particular, we
have $I^-(x) = I^-(y)$ if and only if $I^{-p}(x) =
I^{-p}(y)$. 
 
For a spacetime, the physical significance of
the $\ll^{p}$ relation can be viewed this way: 
Consider any observer as compiling an
encyclopedia of all events witnessed; the spacetime is to be
thought of also as the set of encyclopedias, with the event $p$
corresponding to the history of all events in $I^-(p)$.  Then $p
\ll^{p} q$ means that not only has
$q$ transcribed all the events that are written at $p$, but,
moreover, has had a chance to think about them (i.e., an
observer has traveled from some event $w$, which has recorded all
events that are written at $p$, and then continued on to
event $q$).  

We will use $X^{p}$ to denote $X$ with the relation
$\ll^{p}$; call it the {\it past-determination\/}
of $X$.

If $c = \{x_n\}$ is a future chain in $X^{p}$,
then we have for all $n$, $I^-(x_n)
\subset I^-(w_n)$ for some sequence $\{w_n\}$
with $w_n \ll x_{n+1}$.  This gives us $w_n \in
I^-(x_{n+1}) \subset I^-(w_{n+1})$, so $w_n \ll
w_{n+1}$: $c' = \{w_n\}$ is a future chain in
$X$.  Note that $I^-[c']$ = $I^-[c]$, due
to the interweavings $I^-(x_n) \subset I^-(w_n)$
and $I^-(w_n) \subset I^-(x_{n+1})$; in
general, when this interweaving obtains, we will
say that $c'$ (in $X$) is an {\it associated\/}
chain to $c$ (in $X^{p}$).  This works well for
looking at future limits:

\proclaim{9. Lemma} Let $c'$ be a future chain in
$X$ associated to a future chain $c$ in $X^{p}$. 
Then a point $x$ is a future limit of $c$ in
$X^{p}$ if and only if it is a future limit of $c'$
in $X$.\endproclaim

\demo{Proof} Suppose $I^{-p}(x) = I^{-p}[c]$; then
$I^-(x) =  I^-[c] = I^-[c']$.  Now suppose
$I^-(x) = I^-[c']$; then $I^-(x) = I^-[c]$.  Let
$c = \{x_n\}$.  For $z \ll^{p} x$, we have
$I^-(z)$ is non-empty and
$I^-(z) \subset I^-(w)$ for some $w \ll x$; since
$I^-(x) = I^-[c]$, we have $w \ll x_n$ for some
$n$, so $z \in I^{-p}[c]$.  On the other hand, if
$z \in I^{-p}[c]$, then $I^-(z)$ is
non-empty and $I^-(z) \subset I^-(w)$ for some $w
\ll x_n$ for some $n$; then (since $I^-[c] = I^-(x)$) $w
\ll x$ also, so $z \ll^{p} x$.\qed\enddemo

\proclaim{10. Proposition}  For any
pre-chronological set $X$, $X^{p}$ is a
past-determined pre-chronological set.  If $X$
is a chronological set, so is $X^{p}$; if $X$ is
past-distinguishing, so is $X^{p}$; and if $X$
is future-complete, so is $X^{p}$.\endproclaim 

\demo{Proof} This is largely routine.  For example,
to show $X^{p}$ is past-determined, consider
$I^{-p}(x) \subset I^{-p}(w)$ with
$I^{-p}(x)$ non-empty and $w
\ll^{p} y$; we want to show $x \ll^{p} y$.  From $w
\ll^{p} y$ we know that for some $v
\ll y$, $I^-(w) \subset I^-(v)$ (this is true
even if $w \ll y$).  We have
$I^-(x) \subset I^-(w)$, so $I^-(x) \subset
I^-(v)$.  Since $I^{-p}(x)$ is non-empty, so is
$I^-(x)$.  Thus, $x \ll^{p} y$.

Suppose $X$ is future-complete.  Let $c =
\{x_n\}$ be a future chain in $X^{p}$.  Let $x$
be a future limit of an associated future
chain $c'$ in $X$; then by Lemma 9, $x$ is also a
future limit of $c$ in $X^p$.
\qed\enddemo

Past-determination works well with
future-continuous functions, so long as the
domain has the property that every point is a
future limit of some future chain; let us call this
property being {\it past-connected\/}.  Clearly,
any spacetime is past-connected, and
future completion preserves being past-connected
(since an element of the future chronological
boundary is always a future limit of any of its
generating chains).  Also, forming the
past-determination of a pre-chronological set
preserves being past-connected, in virtue of Lemma
9.

\proclaim{11. Proposition}  If $f: X \to
Y$ is a future-continuous map between
pre-chrono\-logical sets, and $X$ is past-connected,
then the same function $f^{p} = f: X^{p}
\to Y^{p}$ between the past-determinations is also
future-continuous.\endproclaim

\demo{Proof} First we will show that $f^{p}$ is
chronological; we must show that if $x
\ll^{p} y$, then $f(x) \ll^{p} f(y)$:  We have
$I^-(x)$ is non-empty and $I^-(x) \subset I^-(w)$
for some $w \ll y$.  We have some $z \ll
x$, so $f(z) \ll f(x)$: $I^-(f(x))$ is
non-empty.  With $x$ a future limit of a
future chain $c$,
$f(x)$ is a future limit of $f[c]$, so
$I^-(f(x))$ = $I^-[f[c]]$.  We have $c
\subset I^-(w)$, so $f[c] \subset
I^-(f(w))$, so $I^-[f[c]] \subset
I^-(f(w))$.  Therefore, $I^-(f(x)) \subset
I^-(f(w))$ and $f(w) \ll f(y)$: $f(x)
\ll^{p} f(y)$.

Now suppose that $c =\{x_n\}$ is a future chain in
$X^{p}$ with future limit $x$.  For $c'$ an
associated chain in $X$, we know by Lemma 9 that
$x$ is also a future limit in $X$ of $c'$; then
$f(x)$ is a future limit in $Y$ of $f[c']$. 
Furthermore, $f[c']$ is an associated chain of
$f[c]$ (the relations`defining association are
preserved by a chronological function), so Lemma
9 also gives us that $f(x)$ is a future limit in
$Y^{p}$ of $f[c]$. \qed\enddemo

For any pre-chronological set $X$, let us denote
by $\iota^{p}_X : X \to X^{p}$ the map which, on
the set level, is the identity map.  This is
clearly chronological, and it is also
future-continuous:  If, for a future chain $c$
and point $x$ in $X$, $I^-(x) = I^-[c]$, then
$I^{-p}(x) = I^{-p}[c]$ also (using the second
property of $\pl$ mentioned at the beginning of
this section).

\proclaim{12. Corollary} Let $f: X \to
Y$ be a future-continuous map between
pre-chrono\-logical sets, with $X$ past-connected 
and $Y$ past-determined; then there is a unique
future-continuous map $f^{p}: X^{p} \to Y$ such
that $f^{p} \circ \iota^{p}_X = f$.\endproclaim

\demo{Proof} By Proposition 10, $Y^{p} = Y$. 
Then the map $f^{p}: X^{p} \to Y^{p}$ given by
Proposition 11 fills the required condition,
and uniquely does so because of what
$\iota^{p}_X$ is on the set level.\qed\enddemo  

Proposition 11 and Corollary 12 have the same categorical
properties as Proposition 6 and Corollary 7, and we derive
similar categorical constructions: We have a functor
$\text{\bf{p}} :
\text{\bf{PconChron}} \to
\text{\bf{PdetPconChron}}$, where
$\text{\bf{-Pcon-}}$ denotes the subcategory of
past-connected objects.  We also have a natural
transformation
$\boldsymbol\iota^{\text{\bf p}} : 
\text{\bf I} \to
\text{\bf U}^{\text{\bf p}} \circ \text{\bf p}$, where 
$\text{\bf U}^{\text{\bf p}} :
\text{\bf{PdetPconChron}}
\to \text{\bf{PconChron}}$ is the forgetful
functor and
$\text{\bf I}$ is the identity functor on
$\text{\bf{PconChron}}$.  Then Corollary 12 is
precisely the statement that $\text{\bf p}$ (via
$\boldsymbol\iota^{\text{\bf p}})$ is a left
adjoint to $\text{\bf U}^{\text{\bf p}}$.  (The
same applies to the restriction, in all cases,
to the respective subcategories of
past-distinguishing objects, {\bf -Pdis-},
which will be used for composition with the
future completion functor.)  Thus, similar to
Theorem 8, we can conclude that the
past-determination functor is the unique way to
create a past-determined chronological set (at
least, if one starts with all points being
future limits), in a categorical, natural, and
universal manner.  Summarizing as before, we have
\roster

\item[4] For $X$ in {\bf PconPrechron}, $Y$ in {\bf Prechron}, and
$f: X \to Y$ future-continuous, $f^p : X^p \to Y^p$ is the unique
future-continuous map with $f^p \circ \iota^p_X = \iota^p_Y \circ
f$.

\item[5] For $X$ in {\bf PconPrechron}, $Y$  in {\bf PdetPrechron},
and $f: X \to Y$ future-continuous, $f^p : X^p \to Y$ is the
unique future-continuous map with $f^p \circ \iota^p_X = f$.

\item[6] Restricted to {\bf PconChron} (or, more
generally, {\bf PconPrechron}), the
construction of
$X^p$, $f^p$, and $\iota^p_X$ is functorial, natural, and
universal, hence, the unique way to fulfill (4) and (5) in such a
manner.  The same is true for the further
restriction to {\bf PconPdisChron}.
\endroster

We can combine these results with those of the
previous section, composing the categorical items: 
Restricting the functor $\;\widehat{}\;$ to {\bf
PdetPconPdisChron}, we have the composition
$\;\widehat{}\; \circ \text{\bf p}: \text{\bf
PconPdisChron} \to \text{\bf FcplPdetPconPdisChron}$
and the forgetful functor $\text{\bf U}^{\boldkey +} =
\hat{\text{\bf U}} \circ
\text{\bf U}^{\text{\bf p}}$ in the opposite direction.  For any
$X$ in {\bf PconChron}, let $\iota^{\sssize\uparrow}_X
= \hat\iota_{X^p} \circ\iota^p_X : X \to
\widehat{X^p}$; then the collection of maps
$\{\iota^{\sssize\uparrow}_X\}$ forms a natural
transformation
$\boldsymbol\iota^{\sssize\boldsymbol\uparrow} : \text{\bf I}
\;\dot\to\; \text{\bf U}^{\boldkey +}
\circ \,\widehat{}\, \circ \text{\bf p}$ ({\bf I} as above),
and this yields $\;\widehat{}\, \circ \text{\bf p}$ as left
adjoint to $\text{\bf U}^{\boldkey +}$.  Summarizing as
before, we have
\roster

\item[7] For $X$ in {\bf PconChron}, $Y$ in {\bf
PdisChron}, and $f : X \to Y$
future-\linebreak continu\-ous,
$\widehat{f^p} : \widehat{X^p}
\to \widehat{Y^p}$ is the unique future-continuous map with
\linebreak
$\widehat{f^p} \circ \iota^{\sssize\uparrow}_X =
\iota^{\sssize\uparrow}_Y \circ f$.

\item[8] For $X$ in {\bf PconChron}, $Y$ in {\bf
FcplPdetPdisChron}, and $f : X \to Y$ future-continuous,
$\widehat{f^p} : \widehat{X^p}
\to Y$ is the unique future-continuous map with
$\widehat{f^p} \circ \iota^{\sssize\uparrow}_X = f$.

\item[9] Restricted to {\bf PconPdisChron}, the construction of
$\widehat{X^p}$, $\widehat{f^p}$, and $\iota^{\sssize\uparrow}_X$
is functorial, natural, and universal, hence, the unique way to
fulfill (7) and (8) in such a manner.
\endroster
\medpagebreak

We need to consider how past-determination
interacts with future completion.  First, let
us compare the past of a point $x$ of $X$ in
$\hat X^{p}$ with its past in $\widehat{X^{p}}$:
For $y \in X$, we have $y \in \hat I^{-p}(x)$  if
$y \ll x$ or if $\hat I^-(y)$ is non-empty and
$\hat I^-(y) \subset \hat I^-(w)$ for some $w \ll
x$; while we have $y \in \widehat{I^{-p}}(x)$ if
$y \ll x$ or if $I^-(y)$ is non-empty and
$I^-(y) \subset I^-(w)$ for some $w \ll x$. 
These are equivalent conditions:  For instance,
suppose $I^-(y) \subset I^-(w)$; then $P \ll y$
means $P \subset I^-(z)$ for some $z \ll y$,
which then yields $z \ll w$, so $P \ll w$,
also.  For $Q \in \hat\partial(X)$, we have
$Q \in \hat I^{-p}(x)$ if $Q \ll x$---i.e., if $Q
\subset I^-(w)$ for some $w \ll x$---or if
$\hat  I^-(Q)$ is non-empty and
$\hat I^-(Q) \subset \hat I^-(w)$ for some $w \ll
x$; while for $Q \in \hat\partial(X^{p})$, we have
$Q \in \widehat{I^{-p}}(x)$  if $Q \subset
I^{-p}(w)$ for some $w \ll^{p} x$.  But to
compare these two conditions directly is
awkward, and we'll instead use a pair of maps
to aid in the analysis.    

For a chronological set
$X$, let us define the maps $i: {\hat X}^{p} \to
\widehat{X^{p}}$ and
$j: \widehat{X^{p}} \to {\hat X}^{p}$ as follows: 
Both maps are to be the identity on $X$.  For $P
\in \hat\partial(X)$, let $c$ be any future chain
in $X$ generating $P$; then $i(P) = I^{-p}[c]$. 
For $Q \in \hat\partial(X^{p})$, let $c$ be a
future chain in $X^{p}$ generating $Q$; let
$c'$ be an associated chain in $X$, and let
$j(Q) = I^-[c']$.

\proclaim{13. Proposition} For any
chronological set $X$, the maps $i$ and $j$
above are future-continuous and give an
isomorphism between ${\hat X}^{p}$ and
$\widehat{X^{p}}$.\endproclaim

\demo{Proof} We will use $\hl$
to denote the relation in $\hat X$,
$\phl$ to denote the relation in
${\hat X}^{p}$ and
$\hpl$ to denote that in
$\widehat{X^{p}}$.  

First we will treat $i: {\hat
X}^{p} \to \widehat{X^{p}}$.  We have that for $x
\in X\;(= X^{p})$, $i(x) = x$, and for $P \in 
\hat\partial(X)$, $i(P) = I^{-p}[c]$ for a
future chain $c$ in $X$ generating $P$; we
need to see that this is well-defined.  Clearly, 
$I^{-p}[c]$ is an {\eightpoint IP} in $X^{p}$; it is in
$\hat\partial(X^{p})$ because if $I^{-p}[c] =
I^{-p}(x)$, then $I^-[c] = I^-(x)$, which is impossible
with $P \in \hat\partial(X)$.  We can also write it
as $I^{-p}[P]$, which shows that it is
independent of the chain taken as the generator
for $P$.

To show $i$ is chronological, consider $x$ and $y$
in $X$, $P$ and $Q$ in $\hat\partial(X)$, and
future chains $c = \{x_n\}$ and $d = \{y_n\}$
generating $P$ and $Q$ respectively:

It is largely routine to show that 
if $x \phl y$, then  $x \hpl y$.  Since $i(x) = x$
and $i(y) = y$, we are done.

If $x \phl Q$, then we have $x \hl Q$ or we have $\hat I^-(x) \neq
\emptyset$  and $\hat I^-(x) \subset \hat I^-(w)$ for some $w
\hl Q$ (as above, we can take $w \in X$).  In the first case,
we have $x \in Q = I^-[d] \subset I^{-p}[d]$.  In the
second case, we have $I^-(x) \neq \emptyset$ and $I^-(x) \subset
I^-(w)$ and $w \ll y_n$ for some $n$ (since $w \in Q$), so $x \pl
y_n$, so again $x \in I^{-p}[d]$. In either case, this gives us $x
\in i(Q)$, so $i(x) \hpl i(Q)$, and we are done.

If $P \phl y$ (or $Q$), then we have that $P \hl y$ (or $Q$)---in
which case we have $P \subset I^-(w)$ for some $w \ll y$ (or $w
\in Q$)---or we have that $\hat I^-(P) \neq \emptyset$  and
$\hat I^-(P) \subset \hat I^-(w)$ for some $w \hl y$ (or $Q$)---in
which case we again have $P \subset I^-(w)$ and $w \ll y$ (or $w
\in Q$).  Then in either case we have $I^-[c] \subset I^-(w)$,
so $i(P) = I^{-p}[c] \subset I^{-p}(w)$, with $w \pl i(y)$ (or $w
\in i(Q)$), so $i(P) \hpl i(y)$ (or $i(Q)$), and we are done.
\smallpagebreak

Now we consider $j: \widehat{X^{p}} \to {\hat
X}^{p}$.  For $x \in X^{p}$, $j(x) = x$, and for
$Q \in \hat\partial(X^{p})$, $j(Q) = I^-[c']$
for $c'$ a future chain in $X$ associated to a
chain $c$ in $X^{p}$ generating $Q$.  Clearly
$I^-[c']$ is an {\eightpoint IP} in $X$; it is in
$\hat\partial(X)$ because if $x$ is a future
limit in $X$ for $c'$, then it is also a future
limit in $X^{p}$ for $c$, which is impossible for
$Q \in \hat\partial(X^{p})$.  We can also write
it as $I^-[Q]$, which shows it is independent of
the chain taken as the generator for $Q$ and of
the chain in $X$ associated to that one.

To show $j$ is chronological, consider  $x$ and $y$
in $X$, $P$ and $Q$ in $\hat\partial(X^{p})$, and
$c = \{x_n\}$ and $d = \{y_n\}$ future chains in
$X^{p}$ generating, respectively, $P$ and $Q$:

It is routine to show that if $x \hpl y$, then $x
\phl y$.  Thus, $j(x) \phl j(y)$.

If $x \hpl Q$, then $x \in Q$. 
To show $j(x) \phl j(Q)$, we must show
either that
$x \;\hat\ll\; I^-[Q]$ or that $\hat I^-(x) \neq
\emptyset$ and $\hat I^-(x) \subset \hat
I^-(w)$ for some $w\;\hat\ll\; I^-[Q]$,
i.e., $w \in \hat I^-[Q]$.  If
$x \in I^-[Q]$, then we have
$x\;\hat\ll\;\hat I^-[Q]$. Otherwise, the only
way for $x$ to be in $Q$ is for there to be some
$q \in Q$ with $x \ll^{p} q$, i.e., $I^-(x) \neq
\emptyset$ and $I^-(x) \subset I^-(w)$ for
some $w \ll q$; then $\hat I^-(x) \subset \hat
I^-(w)$, also.  Since
$w$ is in $I^-[Q]$, this shows
$j(x)\;\hat\ll^{p}\;j(Q)$.

If $P \;\widehat{\ll^{p}}\; y$ (or $Q$), then
$P \subset I^{-p}(w)$ for some $w \ll^{p} y$ (or
in $Q$).  We then have $j(P) = I^-[P] \subset
I^-[I^{-p}(w)] = I^-(w)$.  Taking the case of
$y$, we need only (since $I^-(w) \neq \emptyset$) look
at  $I^-(w) \subset I^-(z)$ for some $z \ll y$.   We
have $j(P) \subset I^-(z)$ (and $I^-(j(P)) = I^-[P]$
contains points in $P$, so is non-empty), so $j(P)
\;\hat\ll^{p}\; y$.  For the case of $Q$, we
have  $w \in Q$, so for some $q \in Q$, $w
\ll^{p} q$; again, we consider only $I^-(w) \subset
I^-(z)$ for some $z \ll q$.  We
have $j(P) \subset I^-(z)$ with $z \in I^-[Q]$, i.e.,
$z \hl j(Q)$; thus, $j(P) \phl  j(Q)$.
\smallpagebreak

Now we consider the compositions:  For $P \in
\hat\partial(X)$, $j(i(P)) = I^-[I^{-p}[P]]
= I^-[P] = P$.  Thus,
we have $j \circ i = 1_{\hat X^{p}}$.  For $Q \in
\hat\partial(X^{p})$,
$i(j(Q)) = I^{-p}[I^-[Q]]$.  This is clearly
contained in $Q$; we must show it the same as
$Q$.  For $q \in Q$, we know there is some $p
\in Q$ with $q \ll^{p} p$, i.e., either $q \ll p$
(in which case we immediately have $q \in
I^-[I^-[Q]] \subset I^{-p}[I^-[Q]]$) or $I^-(q) \neq
\emptyset$ and $I^-(q) \subset I^-(w)$
for some
$w \ll p$.  In the latter case, we can find $z$
with $w \ll z \ll p$, so that we have $q \ll^{p}
z \ll p$, yielding $q \in I^{-p}[I^-[Q]]$.   This
gives us
$i \circ j = 1_{\widehat{X^{p}}}$.

Now we know that $i$ and $j$ provide an
isomorphism of sets-with-a-relation between $\hat
X^{p}$ and $\widehat{X^{p}}$.  Since future limits
are defined purely in terms of the relations, we
automatically get that both maps are
future-continuous.\qed\enddemo

It is tempting to think of the maps $i_X$ and $j_X$, as
collections for all $X$ in {\bf PconPdisChron}, as
forming a pair of natural transformations $\text{\bf i}
: \text{\bf p} \circ \,\sphat\; \to \;\sphat\, \circ
\text{\bf p}$ and $\text{\bf j} : \;\sphat\, \circ
\text{\bf p} \to \text{\bf p} \circ \,\sphat\;$;
since they are inverses of one another, we would
then have a natural isomorphism between the two
constructions. This would then establish that all
that was done with the construction of
$\widehat{X^p}$ can also be done with the construction
of $\hat X^p$.  Since this latter is the construction
that precisely reproduces the GKP future causal
boundary, this is a desirable outcome.  However,
there is a problem:  One cannot compose the functors
in the order $\text{\bf p} \circ \,\sphat\;$, since
Proposition 6 allows us to conclude that $\hat f :
\hat X \to \hat Y$ is future-continuous (starting
with $f: X \to Y$ future-continuous) only if $Y$ is
already past-determined.  But there is a way around
this difficulty, thereby establishing the categorical
nature of the GKP construction:

We will use $X^+$ to denote $\hat X^p$, for
$X$ any chronological set.  For a future-continuous function $f :
X \to Y$ between chronological sets, we cannot in general
consider $\hat f^p$, since $\hat f$ is not in general
future-continuous unless $Y$ is past-determined.  But
so long as $Y$ is past-distinguishing and $X$ is
past-connected, we can use
$\widehat{f^p} : \widehat{X^p} \to \widehat{Y^p}$, which is
future-continuous; then we can define $f^+ = j_Y \circ
\widehat{f^p} \circ i_X : X^+ \to Y^+$.  Finally, we define
$\iota^+_X = j_X \circ \iota^{\sssize\uparrow}_X: X \to X^+$.

\proclaim{14. Theorem} For any chronological set $X$, $X^+$ is
past-determined and future-complete, and $\iota^+_X : X
\to X^+$ is future-continuous; if $X$ is past-connected
(or, respectively past-distinguishing), then so is
$X^+$.  If
$X$ is past-connected and $Y$ is a past-distinguishing
chronological set, then for any future-continuous map $f:
X \to Y$, $f^+: X^+
\to Y^+$ is the unique future-continuous map such that
$f^+ \circ \iota^+_X = \iota^+_Y \circ f$.  If $Y$ is also
past-determined and future-complete, than $f^+ : X^+ \to Y$ is
the unique future-continuous map such that $f^+ \circ \iota^+_X
= f$. \endproclaim

\demo{Proof}  We need to establish that several diagrams
involving the injections $\iota$ commute.  Some of these follow
from categorical principles when in the category {\bf
Pcon\-Chron}, but others involve the particularities of these
maps (especially when past-connectedness is not assumed).

One noncategorical result is the application of
future-completion using the full strength of Proposition 6: 
The purely categorical use of future completion applies 
only to the category of past-determined, past-distinguishing
chronological sets (as per summary statement (3)), but we need
to apply it to the map $\iota^p_X: X \to X^p$, where it would
be pointless to assume $X$ past-determined.  But Proposition 6
gives us the result anyway, as displayed in Diagram
1:

\midinsert
$$
\CD
X               @>\dsize\iota^p_X>>           X^p             \\
@VV\dsize\hat\iota_XV              @VV\dsize\hat\iota_{X^p}V  \\
\hat X          @>\dsize\widehat{\iota^p_X}>> \widehat{X^p}
\endCD
$$
\botcaption{Diagram 1}composition is $\iota^{\sssize\uparrow}_X$
\endcaption\endinsert

The map $\iota^p_X$ is always
future-continuous, as is the standard future
injection $\hat\iota$, whether for $X$ or for $X^p$.  By
Proposition 6, we can define future completion of $\iota^p_X$,
and have it be future-continuous, even without $X^p$ being
past-distinguishing, so long as for any future chain $c$ in
$X$ generating an element of $\hat\partial(X)$, $\iota^p_X[c]$
has no more than one future-limit in $\widehat{X^p}$; but this
is manifestly the case, since $\hat\partial(X^p)$ (or
$\hat\partial$ in general) is defined in such a way that no two
elements can have identical pasts.  Proposition 6 also yields
that Diagram 1 commutes, and, as indicated, the composition is
$\iota^{\sssize\uparrow}_X$.

For $X$ past-connected, an entirely categorical result comes
from starting with $\hat\iota_X: X \to \hat X$ and
applying the past-determination functor and the natural
transformation
$\boldsymbol\iota^{\text{\bf p}}$, yielding the commutative
diagram of future-continuous maps shown in Diagram 2.

\midinsert
$$
\CD
X               @>\dsize\iota^p_X>>           X^p             \\
@VV\dsize\hat\iota_XV              @VV\dsize(\hat\iota_X)^pV  \\
\hat X          @>\dsize\iota^p_{\hat X}>> \hat X^p
\endCD
$$
\botcaption{Diagram 2}composition is $\iota^+_X$ (see
Diagram 4)
\endcaption\endinsert

Even if $X$ is not past-connected, Diagram 2 still commutes
(since each composition is just inclusion), and
$(\hat\iota_X)^p$ is still future-continuous:  For
chronologicality, we need to show that if $x \pl y$ (in $X$),
then $x \phl y$ (in $\hat X$):  If $I^-(x) \subset I^-(w)$ for
$w \ll y$, then $\hat I^-(x) \subset \hat I^-(w)$ and $w \hl y$.
For future-continuity, we need to show that if $x$ is a future
limit of a future chain $c$ in $X^p$, then $x$ is also a future
limit of $c$ in $\hat X^p$:  If $x$ is a future limit of $c$ in
$X^p$, then by Lemma 9 $x$ is a future limit of an associated
future chain $c'$ in $X$; application of the future-continuous
$\hat\iota_X$ shows that $x$ is a future limit of $c'$ also in
$\hat X$; then, applying Lemma 9 once more (with $c'$ being
associated to $c$ in $\hat X$ as well as in $X$), we have $x$ a
future limit of $c$ in $\hat X^p$.

The identification of the composition in Diagram 2 as
$\iota^+_X$ will follow from Diagram 4.  But before that we
must consider Diagram 3:
This commutes because both $i_X \circ (\hat\iota_X)^p$ and
$\hat\iota_{X^p}$ are just inclusion on the set-level. 
Commuting in the other direction, with $j_X$, then follows
automatically.

\midinsert
$$
\CD
X^p             @=                   X^p                   \\
@VV\dsize(\hat\iota_X)^pV        @VV\dsize\hat\iota_{X^p}V \\
\hat X^p 
@>{\dsize i_X}>{\dsize\underset j_X\to\longleftarrow}>    
\widehat{X^p}
\endCD
$$
\botcaption{Diagram 3}
\endcaption\endinsert

The final result involving commuting injections is shown in
Diagram 4.  To establish the commutativity, we first combine
Diagrams 2 and 3 into Diagram 5 and compare with Diagram 1;
we use the uniqueness property of future completion from
Proposition 6 (summary statement (1)) to see that that the
bottom lines of Diagrams 1 and 5 must be the same, and that is
the content of Diagram 4.  (As before, establishing the result
with $i_X$ automatically yields the result for
$j_X$).  Combining this with Diagram 1 and the definition of
$\iota^+_X$ yields the identification alluded to in Diagram 2. 

\midinsert
$$
\CD
\hat X             @=                   \hat X              \\
@VV\dsize\iota^p_{\hat X}V  @VV\dsize\widehat{\iota^p_X}V \\
\hat X^p 
@>{\dsize i_X}>{\dsize\underset j_X\to\longleftarrow}>    
\widehat{X^p}
\endCD
$$
\botcaption{Diagram 4}
\endcaption\endinsert

\midinsert
$$
\CD
X    @>{\dsize\iota^p_X}>>   X^p  @=  X^p \\
@VV{\dsize \hat\iota_X}V   @VV{\dsize (\hat\iota_X)^p}V
@VV{\dsize \hat\iota_{X^p}}V \\ 
\hat X  @>{\dsize \iota^p_{\hat X}}>>  \hat X^p 
@>{\dsize i_X}>>  \widehat{X^p} 
\endCD
$$     
\botcaption{Diagram 5}Diagrams 2 and 3 combined
\endcaption\endinsert  

Finally, in Diagram 6 we consider the diagram derived from
future-continuous $f: X \to Y$ for $X$ past-connected and
$Y$ past-distinguishing.
The top two portions of Diagram 6 commute due to previously
established results (summary statements (1) and (4)).  The
bottom portion commutes, as that is the definition of $f^+$. 
The overall structure of Diagram 6 can be expressed as Diagram
7.

\midinsert
$$
\CD
X  @>{\dsize f}>>  Y  \\
@VV{\dsize\iota^p_X}V  @VV{\dsize\iota^p_Y}V  \\
X^p  @>{\dsize f^p}>>  Y^p  \\
@VV{\dsize\hat\iota_{X^p}}V  @VV{\dsize\hat\iota_{Y^p}}V  \\
\widehat{X^p}  @>{\dsize\widehat{f^p}}>>  \widehat{Y^p}  \\
@A{\cong}A{\dsize i_X}A  @V{\cong}V{\dsize j_Y}V  \\
\hat X^p = X^+   @>{\dsize f^+}>>  Y^+ = \hat Y^p  \\
\endCD
$$
\botcaption{Diagram 6}
\endcaption\endinsert

\midinsert
$$
\CD
X                      @>{\dsize f}>>   Y                     \\
@VV{\dsize\iota^+_X}V                   @VV{\dsize\iota^+_Y}V \\
X^+                    @>{\dsize f^+}>> Y^+                   \\
\endCD
$$
\botcaption{Diagram 7}overall structure of Diagram 6
\endcaption\endinsert

The map $f^+$ is unique for fulfilling its role in Diagram 7,
since for any other future-continuous map $g : X^+ \to Y^+$
making Diagram 7 commute, $i_Y \circ g \circ j_X$ would fill
the role of $\widehat{f^p}$ in Diagram 6, and the uniqueness
portion of summary statement (4) would show $i_Y \circ g \circ
j_X = \widehat{f^p}$, so $g = f^+$.

If $Y$ is also past-determined and future-complete, then $Y^+ =
Y$ and $\iota^+_Y = 1_Y$, yielding the last statement in the
theorem. \qed\enddemo

Theorem 14 enables us to
define the functor $\boldkey + : \text{\bf
PconPdisChron} \to \mathbreak
\text{\bf FcplPdet\-Pcon\-PdisChron}$, with $X^+ = \hat
X^p$ and $f^+ : X^+ \to Y^+$, even though we cannot
define $f^+$ as $\hat f^p$.  We also
have the natural transformation
$\boldsymbol\iota^{\boldkey +} :
\text{\bf I}
\;\dot\to\; \text{\bf U}^{\boldkey +} \circ \boldkey +$
(where {\bf I} denotes the identity on {\bf
PconPdisChron}) yielding
$\boldkey +$ as left adjoint to $\text{\bf U}^{\boldkey
+}$.  Since a functor can have only one left adjoint,
up to natural isomorphism, there must be a natural
isomorphism, then, between
$\boldkey +$ and $\;\widehat{}\, \circ
\text{\bf p}$.  This is provided, in fact, by
$\text{\bf i} : \boldkey + \;\dot\to\;
\;\widehat{}\, \circ \text{\bf p}$ and $\text{\bf j} :
\;\widehat{}\,
\circ \text{\bf p} \;\dot\to\;
\boldkey +$, whose naturality is a triviality.

Expressing this in summary form as before, we have
\roster

\item[10] For $X$ in {\bf PconChron}, $Y$ in {\bf
PdisChron}, and $f : X \to Y$ future-continuous, $f^+
: X^+ \to Y^+$ is the unique future-continuous map
with $f^+ \circ \iota^+_X = \iota^+_Y \circ f$.

\item[11] For $X$ in {\bf PconChron}, $Y$ in {\bf
FcplPdetPdisChron}, and  $f : X \to Y$
future-continuous, $f^+ : X^+ \to Y$ is the unique
future-continuous map with \linebreak $f^+ \circ
\iota^+_X = f$.

\item[12] Restricted to {\bf PconPdisChron}, the
construction of $X^+$, $f^+$, and $\iota^+_X$ is
functorial, natural, and universal, hence, the unique
way to fulfill (10) and (11).

\endroster

For a spacetime $M$, $\hat M^{p}$ is precisely the
addition to $M$ of the GKP future causal boundary: 
$\hat M$ is identified with the {\eightpoint IP}s of $M$ (the
{\eightpoint TIP}s---{\eightpoint IP}s not of the form
$I^-(x)$ for any point $x$---being
$\hat\partial(M)$), and the
$\ll^{p}$ relation is exactly that as specified by GKP. 
Since any strongly causal spacetime $M$ is in
{\bf PconPdisChron}, any ``reasonable" way
of future-completing $M$---an embedding of $M$ into a 
past-determined, past-distinguishing, future-complete
object---must contain the GKP future boundary in the sense of
Theorem 14: there is a unique future-continuous
extension of the embedding to the GKP future completion of $M$.

Typically, one applies these ideas by starting with a
strongly causal spacetime $M$ and a purported future
completion of it, such as a topological embedding of $M$
into another spacetime
$N$, $j : M \to N$, which preserves the
chronological relations of $M$ and which includes future
endpoints for all the endless timelike curves in $M$. 
The actual object which is the purported future
completion of $M$ is the closure of $j(M)$ in $N$; call
this object $Y$, with a chronology relation inherited from
$N$.  Then, since $M$ is past-connected and $Y$ (typically) is
past-distinguishing, we have, from Theorem 14, $j^+: M^+
\to Y^+$; assuming $Y$ is itself future-complete, we have
$Y^+ = Y^p$.  Then, in particular, $j^+$ maps
$\hat\partial(M)$ to the boundary of $j[M]$ in $N$.  If
we had the good sense to choose the enveloping spacetime
$N$ in such a way that every point in the closure of
$j[M]$ was actually a future endpoint of a timelike curve
in $M$ (mediated by $j$)---or if we restrict $Y$ to be
just such points---then $Y$ is mapped onto by $j^+$, with
the boundary-points of $Y$ being covered by
$j^+[\hat\partial(M)]$.  This places strong restrictions on just
what $Y$ can be.

As an example,
consider a spacetime $M$ which is conformal to $K
\times \Bbb L^1$, where $K$ is any compact
Riemannian manifold (recall $\Bbb L^n$ denotes
Minkowski $n$-space).  $M$ is globally hyperbolic
(any $K \times \{t\}$ is a Cauchy surface), hence,
past-determined.  The future chronological boundary of $M$
consists of a single point $i^+$ which is the future limit of
every endless timelike curve.  (This can be seen by
observing that for any point $p \in M$ and any
future-endless timelike curve $c$, $c$ eventually enters
the future of $p$; this is best noted by looking at
projections into the factors $K$ and $\Bbb L^1$.) 
Thus, for any past-distinguishing, past-determined,
future-complete chronological set $Y$ and any
future-continuous map $f : M \to Y$, $f$ has a
unique future-continuous extension to $M \cup
\{i^+\}$; this means precisely that there is some
point $y^+ \in Y$ such that for every endless
timelike curve $c$ in $M$, $y^+$ is the future limit
of $f \circ c$.  In other words, any ``reasonable"
way of putting a future boundary on $M$ essentially
replicates the GKP procedure, producing $i^+$ and
nothing more (any other points in $Y$ which are
neither in $f[M]$ nor $y^+$ are not the future
limit of anything in $f[M]$ and are in that sense
disconnected from $M$).

More generally, for $M$ any strongly causal
spacetime, let $f: M \to Y$ be any
future-continuous map into a past-distinguishing,
future-complete, chronological set.  Applying the
past-determination functor we have $f^{p} :
M^{p} \to Y^{p}$, a future continuous map into a
past-determined, past-distinguishing,
future-complete chronological set.  Thus, we have
the future extension of $f^{p}$ to $f^+ :
M^+ \to Y^{p}$. Since $\hat\partial(M)$
(the GKP future causal boundary of $M$) contains the
future limit of every future-endless timelike curve
in $M$, the same is true of its image in $Y^{p}$
under $f^+$; assuming that $f$ is an embedding of
$M$ into $Y$, this means that the portion of $Y$
which is directly connected to $f[M]$---i.e., the
points of $Y$ which are future limits of chains
coming from $M$---consists, aside from $f[M]$
itself, of a set future-continuously mapped onto by the GKP
future causal boundary of $M$.

\head
Section IV: Generalized Conditions and Results
\endhead

In a spacetime, for every point $x$, $I^-(x)$ is
an {\eightpoint IP}; this is not necessarily so in a
chronological set $X$.  A future chain $c$ may not
have a future limit because the point $x$
which one might nominate for the future
limit of $c$, has a decomposable past. 
However, this can still play the role of a
future limit in a more generalized sense. 
Let us order by inclusion the {\eightpoint IP}s contained
within $I^-(x)$ (for any $x \in X$); if $P$
is maximal in this ordering, call it a {\it past
component\/} of $x$ (so if $I^-(x)$ is an
{\eightpoint IP}, there is only that one past component). 
Let us call $x$ a {\it generalized future limit
\/} of a future chain $c$ if $I^-[c]$ is a past
component of $x$.  As before, if $x$ is a
generalized future limit of $c$, then it's a
generalized future limit of any sub-chain of
$c$; conversely, if it's a generalized future
limit of some sub-chain of $c$, then it's a
generalized future limit of $c$.  Call $X$ {\it
generalized past-distinguishing\/} if whenever
$x$ and $y$ in $X$ share a past component, $x
= y$ (this is a stronger condition than being
past-distinguishing).  If $X$ is generalized
past-distinguishing, a future chain can have no
more than one generalized future limit.  Call
$X$ {\it generalized future-complete\/} if every
future chain has a generalized future limit (a
weaker condition than being future-complete).  Call a
chronological function
$f: X \to Y$ {\it generalized future-continuous\/} if for every
generalized future limit $x$ of a future chain
$c$ in $X$, $f(x)$ is a generalized future
limit of $f[c]$ (this is neither a stronger nor a
weaker property than being future-continuous); a
composition of generalized future-continuous
functions is generalized future-continuous. 
Finally, call a point $x \in X$ {\it
generalized past-determined\/} if $I^-(x)$ is
non-empty and for any $w$ with $I^-(w)$ containing
some past component of $x$, $x \ll y$ for all $y \gg
w$ (a stronger property than being past-determined);
and $X$ is generalized past-determined if all of its
points are.

These generalized notions are what are needed to
provide the mappings into chronological sets with
decomposable pasts for some of the points; a
typical example would be starting with a spacetime
and embedding it into another manifold to create a
boundary for the spacetime, with some of the
boundary points possibly having decomposable
pasts.  We still get useful information,
such as analogues of Proposition 7, Corollary
12, and Theorem 14---allowing a comparison with
the GKP future causal boundary---so long as the
envelopment obeys these generalized notions:

\proclaim{15. Theorem} For any generalized
future-continuous map $f: X \to Y$ with $X$ a
past-connected chronological set and $Y$ a
generalized past-determined, generalized
past-distinguishing, generalized future-complete
chronological set, there are unique generalized
future-continuous maps $\hat f : \hat X \to Y$, $f^{p}
: X^{p} \to Y$, and $f^+ : X^+ \to Y$
such that $\hat f \circ \hat\iota_X = f^{p} \circ
\iota^{p}_X = f^+ \circ \iota^+_X =
f$.\endproclaim

\demo{Proof} First consider $\hat f$:  For any $P \in
\hat\partial(X)$, with $c$ a future chain
generating $P$, $\hat f[c] = f[c]$ must have a unique
generalized future limit $y \in Y$, so $\hat f(P) = y$
is forced.  For $x \ll P$, with $P$ generated by
$c$, we have $x \in P$, so $\hat f(x) = f(x) \in f[P]
= f[I^-[c]] \subset I^-[f[c]]$, which is a past component
of $\hat f(P)$; that implies $I^-[f[c]]
\subset I^-(\hat f(P))$, so $\hat f(x) \ll
\hat f(P)$. For $P \ll x$, with $P$ generated by $c$, we
have $I^-[c] \subset I^-(w)$ for some $w \ll x$, which
implies $c \subset I^-(w)$ (as in the proof of
Proposition 6); therefore,
$I^-[f[c]] \subset I^-(f(w))$ and $f(w) \ll f(x)$.  With
$\hat f(P)$ the generalized future limit of $f[c]$, we
know that $I^-[f[c]]$ is a past component of $\hat f(P)$;
thus, since $\hat f(P)$ is generalized past-determined, we
have $\hat f(P) \ll f(x) = \hat f(x)$.  For $P \ll Q$, we
interpolate $P \ll x \ll Q$.  Thus, $\hat f$ is
chronological.  Generalized future-continuity
follows from that of $f$, since the only new
future limits are those in $\hat\partial(X)$.

Next consider $f^{p}$:  For $x \ll^{p} y$ in $X$,
we have $I^-(x) \subset I^-(w)$ for some $w \ll
y$.  We also have, since $X$ is past-connected,
that $x$ is a future limit of some future chain
$c$, so $I^-[c] \subset I^-(w)$.  Thus, $I^-[f[c]]
\subset I^-(f(w))$ and $f(w) \ll f(y)$.  Since $f$
is generalized future-continuous, $f(x)$ is the
generalized future limit of $f[c]$, i.e.,
$I^-[f[c]]$ is a past component of $f(x)$.  Since
$f(x)$ is generalized past-determined, this gives
us $f(x) \ll f(y)$.  Therefore, $f^{p}$ is
chronological.  For generalized future-continuity,
the proof in Proposition 11 suffices, once we
obtain an extension of Lemma 9, that
$x$ is a generalized future limit of a future chain
$c$ in $X^{p}$ if and only if $x$ is a generalized
future limit of any associated chain $c'$ in
$X$:   

With $I^-[c']$ a past component of $x$ in $X$,
we need to have $I^{-p}[c]$ a past component of $x$
in $X^{p}$ (the other way being easy).  The proof
of Lemma 9 shows that $I^{-p}[c] \subset
I^{-p}(x)$, we just need to show it a maximal {\eightpoint IP}
with that property.  Suppose $d$ is a chain in
$X^{p}$ with $I^{-p}[c] \subset I^{-p}[d] \subset
I^{-p}(x)$; then  $I^-[c'] \subset I^-[d'] \subset
I^-(x)$, where $d'$ is a chain in $X$ associated to
$d$.  Since $I^-[c']$ is maximal with the property of
being an {\eightpoint IP} contained in the past of $x$, we have
$I^-[c'] = I^-[d']$, and it follows that
$I^{-p}[c] = I^{-p}[d]$. 

Finally, $f^+$ comes from the previous two
constructions, just as in Theorem 14.\qed\enddemo

As an example, let us consider the spacetime $M$
formed from $\Bbb L^2$ by deleting
two timelike half-lines:  $L_0 = \{(0,t) | t \le 0\}$ and
$L_ 2 = \{(2,t) | t \ge 2\}$.  Then $\hat\partial(M)$
consists of these {\eightpoint IP}s (illustrated in figure 1):
for each $t \le 0$,
$P_t^L$, bounded by $L_0$ and the null line from $(0,t)$
going to the left in the past, and $P_t^R$, bounded
by $L_0$ and the null line from $(0,t)$ going to
the right in the past; for each
$t > 2$, $Q_t^L$, bounded by $L_2$, the null line
from $(2,t)$ going to the left in the past, and the
null line from $(2,2)$ going to the right in the
past, and $Q_t^R$, bounded by $L_2$, the null line
from $(2,t)$ going to the right in the past, 
the null line going from $(2,2)$ to $(0,0)$, and $L_0$;
$Q_2$, bounded by the null line
from $(2,2)$ going to the right in the past, 
the null line going from $(2,2)$ to $(0,0)$, and $L_0$;
and the usual future causal boundary  for $\Bbb L^2$,
consisting of the pasts of null lines going out to future
infinity, as well as the entire spacetime.  In $\hat M$,
we have $P_0^R \ll (3,4)$ (because $\hat I^-(P_0^R)
\subset I^-(3,3)$), but there is no relation between
$P_0^L$ and $(3,4)$.  Also, since there is no
relation between $(1,1)$ and $(3,4)$, the space
is not past-determined (same for $M$).  The
past-determination of $\hat M$ adds some
relations:  $(s,s) \ll^p (3,4)$ for $0 < s < 2$;
however, even in $\hat M^p$, there is no relation
between $P_0^L$ (or $(x,t)$ for $x<0$ and $t<x$) and
$(3,4)$.

A typical embedding of $M$ to create a boundary for
it would be to use $\hat M$, except to coalesce
$P_0^R$ and $P_0^L$ into a point $P_0$, with $P_0$
not related to $(3,4)$; call this $\bar M$.  This is
not past-determined, since $(1,1)$ is not related to
$(3,4)$.  We can add relations to $\bar M$ by
saying that two points are related if there is a
curve between them, causal in $\Bbb L^2$,
somewhere timelike, allowed to include ``one side"
of $L_0$ or $L_2$ but not to cross either of them
(the points $P$ and $Q$ being identified with one
side or the other of those two lines).  This
space, $\Bar{\Bar M}$, is past-determined,
past-distinguishing, and generalized
future-complete (since $P_0$ has two past
components, it is not future-complete); hence we
get a generalized future-continuous map from $M^+$ to $\Bar{\Bar
M}$; this is just the map coalescing $P_0^R$ and $P_0^L$ into
$P_0$.  Similarly, any means of adding complete boundaries
to the slits in $M$ will end up being a quotient of
the boundaries from $\hat M$.

\head
Appendix: An Example with the Full Causal Boundary
\endhead

Why has the foregoing structure been involved solely with the
future chronological boundary and not with some combination of
future and past, such as the complete causal boundary of the
GKP construction?  In a nutshell, the reason is that
the combined causal boundary, future with past, is, in
general, neither future- nor past-complete:  It's as
if one were to attempt a completion of a Riemannian
manifold by adding endpoints to all finite-length
endless curves, and then one found, for instance, 
that the added endpoints themselves constituted a
finite-length endless curve.  Here is an example of
how that can happen with the GKP causal boundary and a
rather ordinary spacetime (illustrated in figure 2):

Our spacetime $M$ will be a subset of Minkowski 2-space, $\Bbb
L^2$, in which we will cut an infinite number of slits and poke an
infinite collection of holes.  First we select three parallel null
lines as reference objects, say $L^+ = \{t = x+1\}$, $L^0 = \{t =
x\}$, and $L^- = \{t = x-1\}$; we also select a future-timelike
curve $c$ asymptotic (in the future) to the middle null line, say
$c(s) = (\sinh s,\cosh s)$.  For each positive integer $k$,
consider a short null segment centered at $c(2k)$, parallel to
the reference lines, say from $(\sinh(2k) - .5, \cosh(2k) - .5)$
to $(\sinh(2k) + .5, \cosh(2k) + .5)$; denote the past and future
endpoints of these null segments by, respectively, $p_k$ and
$q_k$.  To form $M$, delete from $\Bbb L^2$ each vertical segment
from $p_k$ down to $L^-$, each vertical segment from
$q_k$ up to $L^+$, and each point $c(n)$, all $k$ and all $n$. 

The future causal boundary for $M$ includes an {\eightpoint IP}
$P_n$  corresponding to each of the deleted points
$c(n)$, while the past causal
boundary includes an {\eightpoint IF} $F_n$ for the same. (The
future causal boundary also has, for each of the slits,
{\eightpoint IP}s for each point of the left and right sides of
the slit, except a single {\eightpoint IP} for the bottom point; an {\eightpoint IP} for
future timelike infinity,
$i^+$; and {\eightpoint IP}s for future null infinity, $\Im^+$---but with points corresponding to the null lines between
$L^-$ and
$L^+$ missing from the usual $\Im^+$ for $\Bbb
L^2$, due to the slits.  The past causal boundary also has, for
each slit, {\eightpoint IF}s for each point of the left and right
sides, except a single {\eightpoint IF} for the top; and past
timelike and null infinity,
$i^-$ and $\Im^-$.  However, we shall be largely unconcerned
with these other points in the causal boundary.)  We have each
$P_{2k} \subset P_{2k+1}$, so that within $\hat M$, $P_{2k} \ll
P_{2k+1}$, but no other relations obtain among these boundary
points, due to the slits; and the future chronological completion
of $M$ is precisely the same as adding the GKP future
causal boundary to $M$.  Similarly, within
$\check M$, the past chronological completion of $M$ (the same as
adding the GKP past causal boundary), we have $F_{2k-1} \ll
F_{2k}$ (since $F_{2k-1} \supset F_{2k})$ but no other relations
among those particular boundary points.

There is more than one scheme for combining future and past
causal boundaries into a single causal boundary (see, for
instance, \cite{S1} and \cite{S2} for a good alternative to the
GKP prescription), but they all agree in a spacetime as simple as
this one:  Each $P_n$ is identified with $F_n$, which we'll
call $z_n$ (and the two copies of the left side of each slit---the
one from the future causal boundary and the one from the past
causal boundary---are identified, and the same with the right
sides, with a single boundary point for the top and one for the
bottom of each slit); this is called $M^*$ in \cite{HE}.  Thus,
in any choice of chronology relation on $M^*$ which makes the
obvious mappings from $\hat M$ and $\check M$ to $M^*$
chronological, we must have $z_{2k-1} \ll z_{2k}$ (from $\check
M$) and $z_{2k} \ll z_{2k+1}$ (from $\hat M$), so that $\{z_n\}$
is a future chain in $M^*$.  However, this chain has no future
limit in $M^*$---for instance, the past of $L^0$ is not an {\eightpoint IP},
since the slits decompose that past into an infinite number of
past sets:  In essence, the chain $\{z_n\}$ ``escapes" our
intended completion of $M$ by aiming for the ``hole" in $\Im^+$ left by the missing points corresponding to the null lines
between $L^-$ and $L^+$.  

Szabados, in section 6 of \cite{S1},
proposes a chronology relation on $M^*$ that, in this case, would
leave no relations at all among the $\{z_n\}$; this construction
has the advantage of not adding any additional relations among the
points of $M$---unlike the GKP construction in the future or
past causal boundaries, or the future and past completions
defined here---but the disadvantage of not preserving relations
already defined in $\hat M$ or $\check M$.

\enddocument
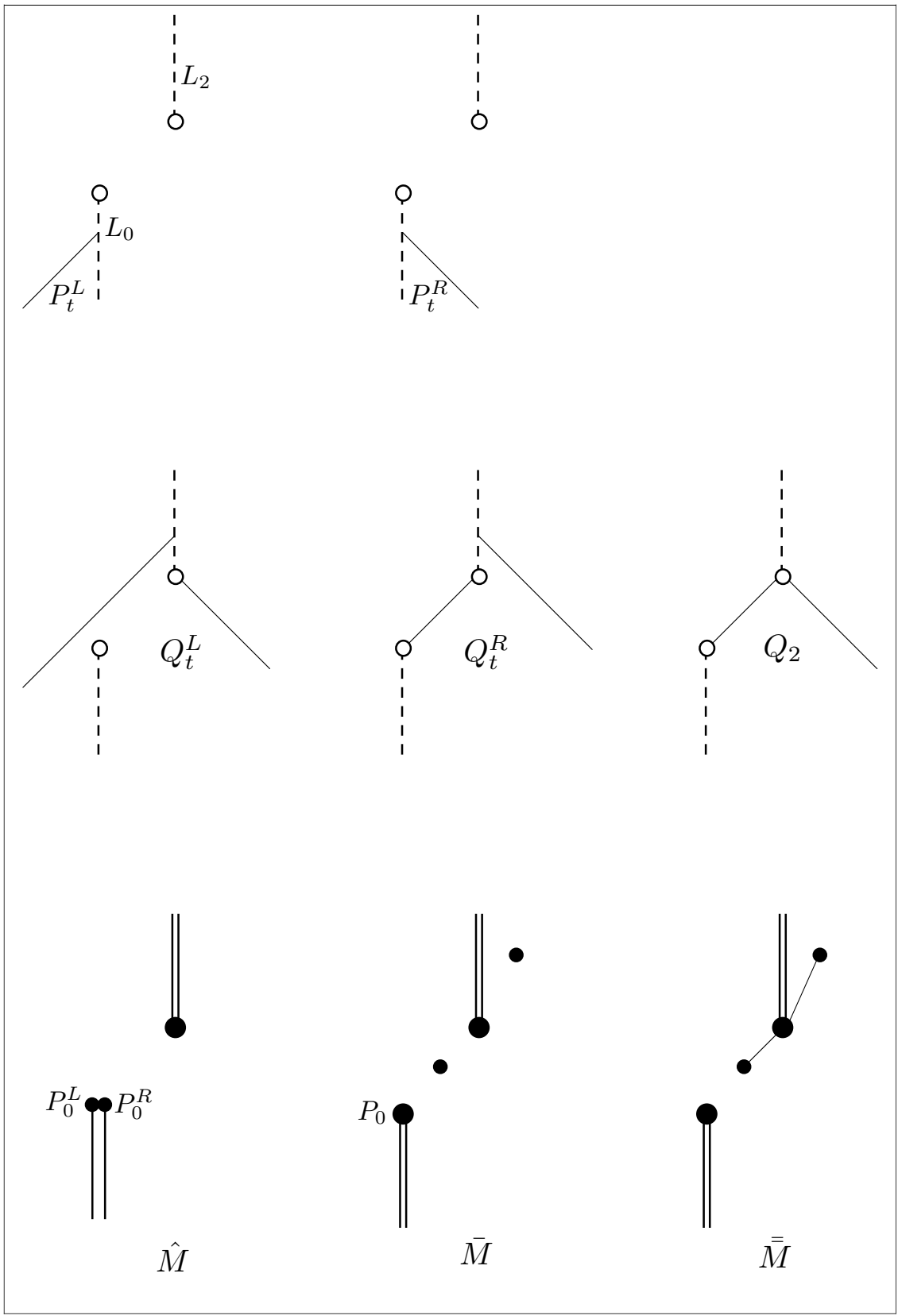
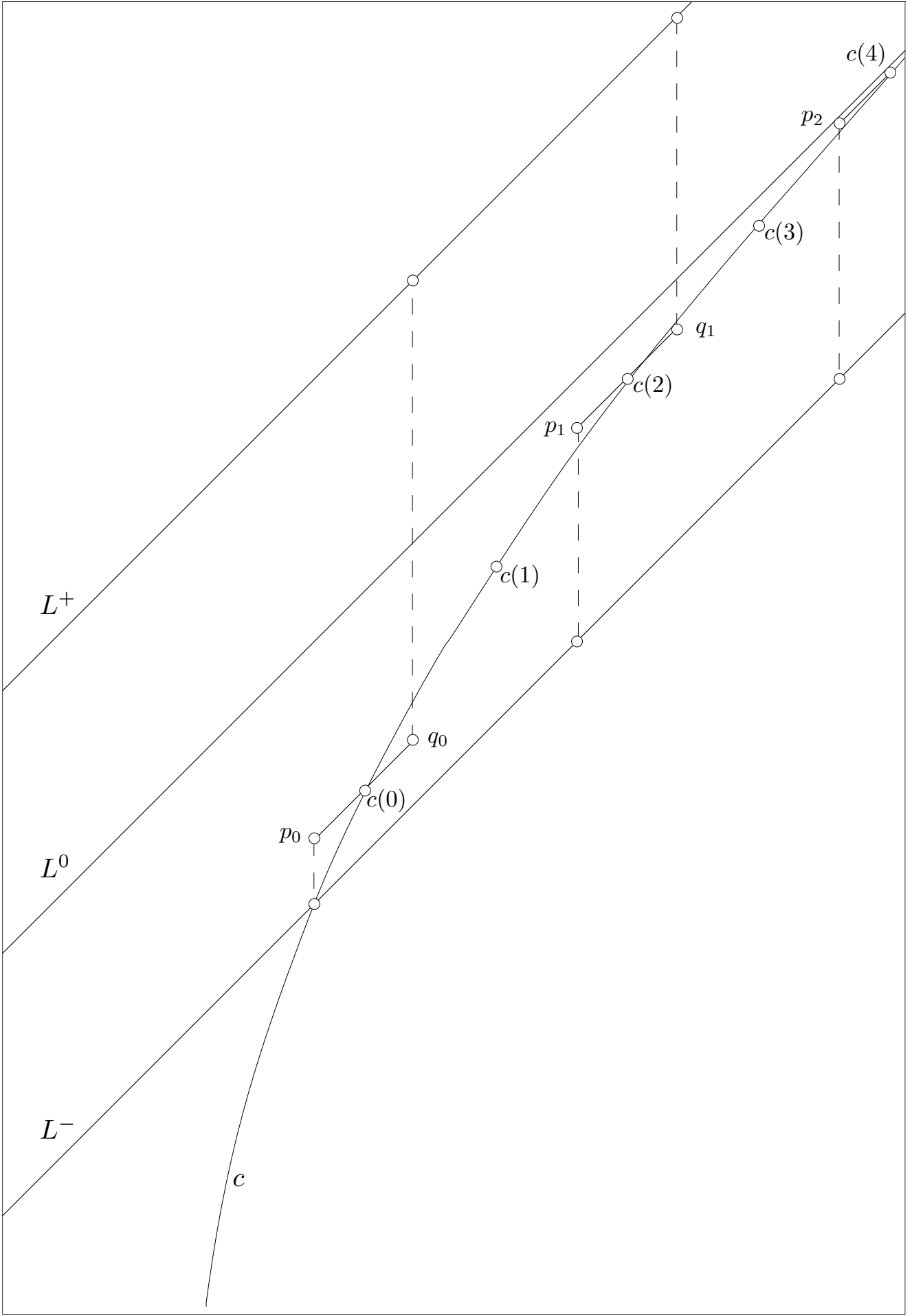
